\documentclass[aps,pre,twocolumn,superscriptaddress]{revtex4-1}

\usepackage{amsmath}
\usepackage{amsfonts}
\usepackage{amssymb}
\usepackage{amsthm}
\usepackage{mathtools}

\usepackage{graphicx}
\usepackage{color}

\usepackage{epsf}
\usepackage{graphicx}
\usepackage{amssymb}

\usepackage{color}

\begin{document}

\title{Superconducting spin valve effect and triplet superconductivity in CoO$_x$/Fe1/Cu/Fe2/Cu/Pb multilayer}

\author{P.~V.\ Leksin}
\affiliation{Leibniz Institute for Solid State and Materials
Research IFW Dresden, D-01171 Dresden, Germany}
\affiliation{Zavoisky Physical-Technical Institute, Russian
Academy of Sciences, 420029 Kazan, Russia}

\author{N.~N.\ Garif'yanov}
\affiliation{Zavoisky Physical-Technical Institute, Russian
Academy of Sciences, 420029 Kazan, Russia}

\author{A.~A.\ Kamashev}
\affiliation{Zavoisky Physical-Technical Institute, Russian
Academy of Sciences, 420029 Kazan, Russia}

\author{Ya.~V.\ Fominov}
\affiliation{L.~D.\ Landau Institute for Theoretical Physics, Russian
Academy of Sciences, 142432 Chernogolovka, Russia}
\affiliation{Moscow Institute of Physics and Technology, 141700
Dolgoprudny, Russia}

\author{J.\ Schumann}
\affiliation{Leibniz Institute for Solid State and Materials
Research IFW Dresden, D-01171 Dresden, Germany}

\author{C.\ Hess}
\affiliation{Leibniz Institute for Solid State and Materials
Research IFW Dresden, D-01171 Dresden, Germany}

\author{V.\ Kataev}
\affiliation{Leibniz Institute for Solid State and Materials
Research IFW Dresden, D-01171 Dresden, Germany}

\author{B.\ B\"{u}chner}
\affiliation{Leibniz Institute for Solid State and Materials
Research IFW Dresden, D-01171 Dresden, Germany}
\affiliation{Institut f\"{u}r Festk\"{o}rperphysik, Technische
Universit\"{a}t Dresden, D-01062 Dresden, Germany}

\author{I.~A.\ Garifullin}
\affiliation{Zavoisky Physical-Technical Institute, Russian
Academy of Sciences, 420029 Kazan, Russia}

\date{\today}

\begin{abstract}
We report magnetic and superconducting properties of the modified
spin valve system CoO$_x$/Fe1/Cu/Fe2/Cu/Pb. Introduction of a Cu
interlayer between Fe2 and Pb layers prevents material
interdiffusion process, increases the Fe2/Pb interface
transparency, stabilizes and enhances properties of the system.
This allowed us to perform a comprehensive study of such
heterostructures and to present theoretical description of the
superconducting spin valve effect and of the manifestation of the
long-range triplet component of the superconducting condensate.

\end{abstract}



\maketitle

\section{INTRODUCTION}

The antagonism of superconductivity (S) and ferromagnetism (F)
arises from the fact that ferromagnetism requires parallel (P) and
superconductivity requires antiparallel (AP) orientation of spins.
The exchange splitting of the conduction band in strong ferromagnets
which tends to align electron spins parallel is larger by orders of
magnitude than the coupling energy for the AP alignment of the
electron spins in the Cooper pairs in conventional superconductors.
Therefore the singlet pairs with AP spins of electrons are easily
destroyed by the exchange field in a homogeneous sample.

Just bearing in mind this strong antagonism, the spin valve effect
in the F1/F2/S multilayer structure, where F1 and F2 are the F
layers with uncoupled magnetizations, and S is the superconducting
layer, has been theoretically predicted \cite{Oh}. Preliminary
analysis showed that at P orientation of magnetizations of F1 and
F2 layers the superconducting transition temperature
$T_c^\mathrm{P}$ is lower than that in the case of their AP
orientation, $T_c^\mathrm{AP}$.

Recently we have experimentally demonstrated a full switching
between the normal and superconducting state in an S/F thin film
heterostructure \cite{Leksin1}. For that we have adapted the
design by Oh {\it et al.} \cite{Oh} and have optimized the
materials choice and the specific geometry of the F1/F2/S scheme.
For this layer sequence, we have chosen the following set of
materials: cobalt oxide for the insulating antiferromagnetic bias
layer which pins the magnetization of the F1 layer; Fe for the
ferromagnetic F1 and F2 layers; Cu as a normal (N) metal layer
between the F1 and F2 layers to decouple their magnetizations, and
finally In for the S layer.

Our further important result was the observation of the
sign-changing behavior of the spin valve effect $\Delta
T_c=T_c^\mathrm{AP}-T_c^\mathrm{P}$ by varying the F2-layer
thickness $d_\mathrm{Fe2}$ \cite{Leksin2,Leksin3}. To understand
this observation, we have analyzed our results in the framework of
the theory by Fominov {\it et al.} \cite{Fominov2} which predicts
an oscillating behavior of $\Delta T_c$ due to the interference
effects for the superconducting pairing function reflected from
both surfaces of the F2 layer.

This theoretical paper generalizes the theory of the spin valve
effect for F1/F2/S structure taking into account the appearance of
the long-range triplet component (LRTC) in the superconducting
condensate predicted by Bergeret {\it et al.} \cite{Bergeret1}.
According to Ref. \cite{Fominov2} the appearance of the triplet
component in the reference F1/F2/S spin valve system should manifest
itself as a minimum in the in-plane angular dependence of $T_c$ at a
noncollinear orientation of magnetizations of the F1 and F2 layers.
Unfortunately, such experiment for the CoO$_x$/Fe1/Cu/Fe2/In system
turned out to be unrealizable under the well-controlled conditions
owing to a low value of $T_c$ for indium and its extreme sensitivity
to small out-of-plane tilting of the external magnetic field. In
this respect lead has much better superconducting critical
parameters which have determined its choice as an S layer in our
previous study \cite{Leksin4}. Indeed, in the multilayer spin-valve
heterostructure CoO$_x$/Fe1/Cu/Fe2/Pb we have observed indications
of the LRTC in the superconducting condensate arising due to the
penetration of the Cooper pairs into two F layers with noncollinear
orientation of magnetizations. The LRTC manifested in a nonmonotonic
behavior of the superconducting transition temperature $T_c$ of the
Pb layer upon gradual rotation of the magnetization of the
ferromagnetic F2 layer $\boldsymbol M_{\rm Fe2}$ with respect to the
magnetization of the F1 layer $\boldsymbol M_{\rm Fe1}$ from the P
to the AP orientation. We observed a clear minimum of $T_c$ for a
noncollinear configuration of $\boldsymbol M_{\rm Fe1}$ and
$\boldsymbol M_{\rm Fe2}$. As follows from our analysis in the
framework of the theory of the superconducting triplet spin valve
\cite{Fominov2}, such minimum of $T_c$ is a fingerprint of the LRTC
generated by noncollinear magnetizations $\boldsymbol M_{\rm Fe1}$
and $\boldsymbol M_{\rm Fe2}$.

Though we could achieve qualitative understanding of the
experimental signatures of the LRTC, a rapid degradation of our
heterostructures and a nonperfect properties of the F2/S interface
prevented to obtain a comprehensive picture of the generation and
experimental manifestation of the triplet component in the
superconducting condensate in the S/F spin valve.

The major goal of the present work was twofold: (i) to improve the
stability and the performance of the S/F heterostructures; (ii) to
investigate the physical properties of the modified structures
and to provide a theoretical description of the ordinary spin valve
and the LRTC effects.

The paper is organized as follows. For the sake of completeness we
briefly review the spin valve effect and the generation of LRTC in
spin valve structures in Section~II. The preparation technique and
optimization of the properties of the films is described in
Section~III. In Section IV we present magnetic and normal state
transport properties. Section V contains the main experimental
data on the superconducting and spin valve properties of the
samples. In Section VI we perform  theoretical analysis of the
obtained results. Finally, the work is summarized in Section VII.

\section{State of art}
\subsection{Superconductor/ferromagnet proximity effect}

 In a metallic ferromagnet there exists the $sd$-exchange interaction
between spins of localized moments ($S$) and spins of conduction
electrons ($s$) in the form of $\hat H_\mathrm{ex}= \sum_i
J_{sd}(\mathbf S_i \mathbf s) \delta(\mathbf r-\mathbf r_i)$. If
localized moments with concentration $n_S$ are aligned
ferromagnetically (in parallel), then the spatially averaged
Hamiltonian becomes $\hat H_\mathrm{ex}= n_S J_{sd}(\mathbf S
\mathbf s)$. This ferromagnetic interaction means that conduction
electrons with spin-up and spin-down orientations have different
energies. Thus, the conduction band in the F layer turns out to be
split into two subbands. The splitting (the difference of energies
for spin up and spin down) is then $n_S J_{sd} S$. For {\it
3d}-transition metal ferromagnets (such as Mn, Fe, or Co) this
splitting is of the order of 1 eV. One can imagine that this is a
result of the Zeeman splitting under the influence of the
so-called exchange field $h$. In terms of the exchange energy, the
splitting is $2h$.

In turn, the coupling energy for the electrons in the Cooper pairs
according to the Bardeen-Cooper-Schrieffer (BCS) microscopic
theory of superconductivity amounts to $\Delta =1.76 k_B T_c$
\cite{BCS} (here $k_B$ is the Boltzmann constant and $T_c$ is the
superconducting transition temperature). It is three orders of
magnitude smaller than the splitting of the conduction band  of a
{\it 3d} ferromagnet. That is why {\it 3d} ferromagnetism strongly
suppresses conventional singlet superconductivity. For this reason
in thin film superconductor/ferromagnet heterostructures the
Cooper pairs can penetrate into an F layer over a small distance
$\xi_h$ only. Considering the superconducting layer in the
``dirty'' limit, i.e. in the regime when the mean free path of
conduction electrons in a superconductor $l_s$ is much smaller
compared to the BCS coherence length $\xi_0$, the characteristic
depth of the decay of the pairing function in the F layer
$\xi_h=(\hbar D_F/h)^{1/2}$ is given by the diffusion coefficient
$D_F$ and the exchange energy $h$ of the F layer \cite{Radovic}.
For iron film the value of $\xi_h$ is less than 0.8 nm (see, e.g.,
\cite{Lazar}).

\subsection{Superconducting spin valve}

The physical origin of the spin valve effect based on the S/F
proximity effect relies on the idea to control the pair-breaking,
and hence the superconducting transition temperature $T_c$, by
manipulating the mutual orientation of the magnetizations of the F
layers in a heterostructure comprising, e.g., two F and one S
layer in a certain combination. The first spin valve structure
based on the S/F proximity effect has been proposed by Oh {\it et
al.} \cite{Oh}. According to this theory the superconducting spin
valve effect in F1/F2/S structures takes place because the mean
exchange field from two F layers acting on Cooper pairs in the S
layer is smaller for the AP orientation of the magnetizations of
these F layers compared to the P case. However, it has been shown
recently \cite{Fominov1,Fominov2} that the situation is not that
simple. It is well known (see, e.g., the review \cite{buzdin})
that in a ferromagnetic layer the Cooper pair acquires a nonzero
momentum due to the Zeeman splitting of the electron levels and
thus its wave function oscillates in space. If the F layer is
sufficiently thin, the wave function reflected from the surface of
the F layer opposite to the S/F interface can interfere with the
incoming one. Depending on the F layer's thickness, the
interference at the S/F interface may be constructive or
destructive. This should apparently lead to an enhancement of
$T_c$ of S/F structure (direct spin valve effect with
$T_c^\mathrm{AP}>T_c^\mathrm{P}$ in a spin valve structure) or to
its decrease (inverse spin valve effect,
$T_c^\mathrm{AP}<T_c^\mathrm{P}$), respectively. Another structure
proposed also theoretically \cite{Tagirov,Buzdin,Buzdin1} is the
scheme F1/S/F2  with the operation principle similar to that
described above. For both structures the interference effects can
lead to nontrivial behavior of $T_c$. At the same time, the
difference in operation of these two structures consist in that
for F1/F2/S structure the direct and inverse spin valve effects
can be realized and for F1/S/F2 structure only the direct effect
is expected.

A large number of works confirmed the predicted influence of the
mutual orientation of the magnetizations in the F1/S/F2 structure on
$T_c$ (see, e.g., \cite{Gu,You,Potenza,Pena,Moraru,Miao}). However,
the magnitude of the spin valve effect $\Delta
T_c=T_c^\mathrm{AP}-T_c^\mathrm{P}$  turns out to be smaller than
the width of the superconducting transition $\delta T_c$ itself.
Hence a full switching between the normal and the superconducting
state was not achieved. Structures similar to suggested by Oh {\it
et al.} \cite{Oh} were investigated to a less extent. A studied
[Fe/V]$_n$ antiferromagnetically coupled superlattice with thick
superconducting vanadium layer on the top instead of a single
F1/N/F2 trilayer \cite{Westerholt,Nowak,Nowak1} was not actually the
spin valve device because the system could not be switched from the
AP to P orientation of the magnetizations instantaneously. At the
same time the analysis of the temperature dependence of the critical
field has shown that implicitly $\Delta T_c$ of this structure can
reach up to 200 mK at $\delta T_c \sim$100 mK.

Comparison of the experimental results obtained for both proposed
structures of the spin switches gives grounds to conclude that the
scheme proposed by Oh {\it et al.} \cite{Oh} may be the most
promising for the realization of the full spin switch effect for
the superconducting current in an S/F thin film heterostructure.
Later on a set of asymmetric structures  was theoretically
proposed \cite{proshin,avdeev,avdeev1} which are not yet
experimentally tested.

\subsection{Long-range triplet superconductivity}

Bergeret {\it et al.} \cite{Bergeret1} underlined the fact that
the odd-frequency triplet superconductivity arises in S/F systems
with conventional \textit{s}-wave singlet BCS-type
superconductors; they further demonstrated that in the case of
inhomogeneous magnetization $\boldsymbol M_{\rm F}$ of the F part,
some of the odd-frequency superconducting components become
long-range. The diffusive Josephson junctions with the F layers
and noncollinear $\boldsymbol M_{\rm F}$ orientation where
long-range Josephson coupling may exist, have been analyzed in
Refs. \cite{bergeret2,houzet,eschrig}.
 Volkov {\it et al.} \cite{bergeret2}
considered F$'$SFSF$'$ structure with different magnetization
orientations in the F and F$'$ layers. It was shown that in the
F$'$ layer the triplet component arises with the total spin of the
Cooper pair $S_z=\pm 1$. This component is insensitive to the
exchange field and can penetrate into the F layer over a distance
much larger than $\xi_h$. This is the reason why the
superconducting odd $S_z=\pm 1$ component has been named
long-range triplet component. The LRTC can also be generated in
the system with spatial or momentum dependence of the exchange
field, as well \cite{Melnikov,Booniad}.

In order to comprehend this fact, let us assume that the
magnetization of the F2 layer in the F1/F2/S system is directed
along the $y$-axis. When the singlet Cooper pairs enter this
ferromagnetic layer, the triplet component with zero spin
projection on the $y$ axis is generated. Then these Cooper pairs
penetrate into the F1 layer where the magnetization is directed
along $z$ axis. Now the triplet component with zero projection on
the $y$ axis  has projections $\pm 1$ on the $z$ axis, which
corresponds to the equal-spin pairing which is just the LRTC. The
series of experiments were performed (see, e.g., reviews
\cite{bergeret,buzdin,efetov,efetov2}) which show an anomalously
deep penetration of the superconducting condensate into the F
layer typical for the triplet superconductivity.

Finally, we mention that earlier indications for long-range
superconductivity in an F layer have been detected through the
proximity-induced conductance \cite{giroud,petrashov} even before
the theoretical works have appeared. Recently the occurrence of the
odd in the Matsubara frequency triplet superconductivity in the S/F
systems, predicted in Ref. \cite{bergeret2} was inferred from the
experiments on the differential conductance \cite{visani} and on
Josephson junctions through observation of an anomalously deep
penetration of the Cooper condensate into the F layer (see, e.g.,
\cite{sosnin,keizer,chan,blamire,aarts1,khire,Westerholt1}).

We note that our experiments \cite{Leksin4} were advantageous in
that they address the primary superconducting  parameter of the spin
valve, the behavior of $T_c$, which is directly affected by the
spin-triplet component. Recently  for the structure of the spin
valve similar to the studied by us an evidence for generation of
LRTC was obtained as well \cite{zdravkov,Wang}.

\section{Samples}

In our previous study of the spin valve construction
CoO$_x$/Fe1/Cu/Fe2/Pb \cite{Leksin4} the sample degradation on a
time scale of a week was observed. This prevented a long-term
detailed study of the properties of the samples. We suppose that
this degradation was caused by the interdiffusion of Fe and Pb
atoms through the Fe2/Pb interface. To prevent such an
interdiffusion an introduction of the antidiffusion (AD)
interlayer between Fe2 and Pb layers may be helpful. For that
purpose we have chosen a copper film \cite{Leksin5}.

\subsection{Design}

The final design of the samples is depicted in Fig.~1. The
\begin{figure}[h]
\includegraphics[width=0.6\columnwidth]{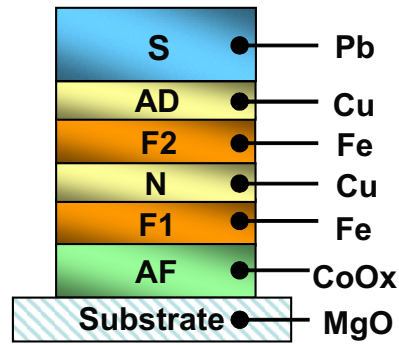}
\caption{(Color online) Design of the samples.}
\end{figure}
CoO$_x$/Fe1 is a standard combination providing a high pinning
effect of the Fe1 layer magnetization. The antiferromagnetic
CoO$_x$ layer has the N\'eel temperature $T_N\sim 250-290$ K,
depending on the oxygen content $x$. After the in-plane magnetic
field cooling procedure from 300 K down to 4 K the magnetization
of the Fe layer contacting the CoO$_x$ layer is strongly pinned in
the direction of the field. However, the Fe2 layer's magnetization
is not pinned since it is decoupled from Fe1 magnetization by a
nonmagnetic 4 nm thick Cu layer. Thus the mutual orientation of
the Fe1 and Fe2 layer's magnetizations can be changed using a
small external magnetic field applied in the desired direction. A
thin Cu layer separates the Fe2 and S layer. As it will be shown
below (see, e.g., Fig.~3) this conducting Cu interlayer does not
influence $T_c$ and the proximity effect between Fe2 and Pb
layers. Since the top Pb layer is air-sensitive all samples were
capped by a 85 nm thick isolating Si$_3$N$_4$ layer. It does not
cause an additional proximity effect with the S layers since it is
not conducting.

\subsection{Preparation}

Metallic layers were grown on high-quality MgO (100) substrates
using classical e-gun technology in ultra-high vacuum (UHV)
conditions of about $1 \cdot 10^{-8}$ mbar within a closed vacuum
cycle. The evaporation chamber has a load-lock station allowing to
avoid vacuum breaking before substrate load. Substrate temperature
was kept at 300 K to avoid undesired alloying between the
neighboring layers. The thickness of the films during the growth
was controlled by a standard quartz-crystal monitor system. All
materials used for evaporation had a purity of better than 4N,
i.e. the contamination level could be kept below 0.01 at \%. The
substrates were fixed at a small rotating wheel on the sample
holder. After that the sample holder was placed inside the
load-lock station. The rotating wheel system allows to prepare a
set of samples with varied parameters at the same evacuation
cycle. The Co oxide layer was prepared in two stages. First, the
metallic Co film was deposited on the substrates. On the second
stage it was transferred to the load-lock station were the Co film
was oxidized using 2 hours exposure at 100 mbar of pure oxygen
gas. After the oxidation procedure the samples were returned to
the UHV deposition chamber where the layers growth process was
continued. On the final stage the samples were transferred to the
neighboring sputtering chamber were they were capped by a 85 nm
thick Si$_3$N$_4$ insulating protection layer. In order to prepare
high quality samples the deposition rates for the materials were
optimized. For Co, Fe, Cu we used the rates of 0.05 nm/s and for
Pb --- 4 nm/s.

\subsection{Fe/Pb interface quality}

Our first studies of the proximity effect in the Fe/Pb samples
revealed that the time stability of the interface between Fe and
Pb materials was rather limited \cite{Leksin5}. Hence the sample
properties were stable within one week period only. Fig.~2 shows
the time evolution of the superconducting properties of the Fe/Pb
systems. In the
\begin{figure}[h]
\includegraphics[height=0.8\columnwidth,angle=-90]{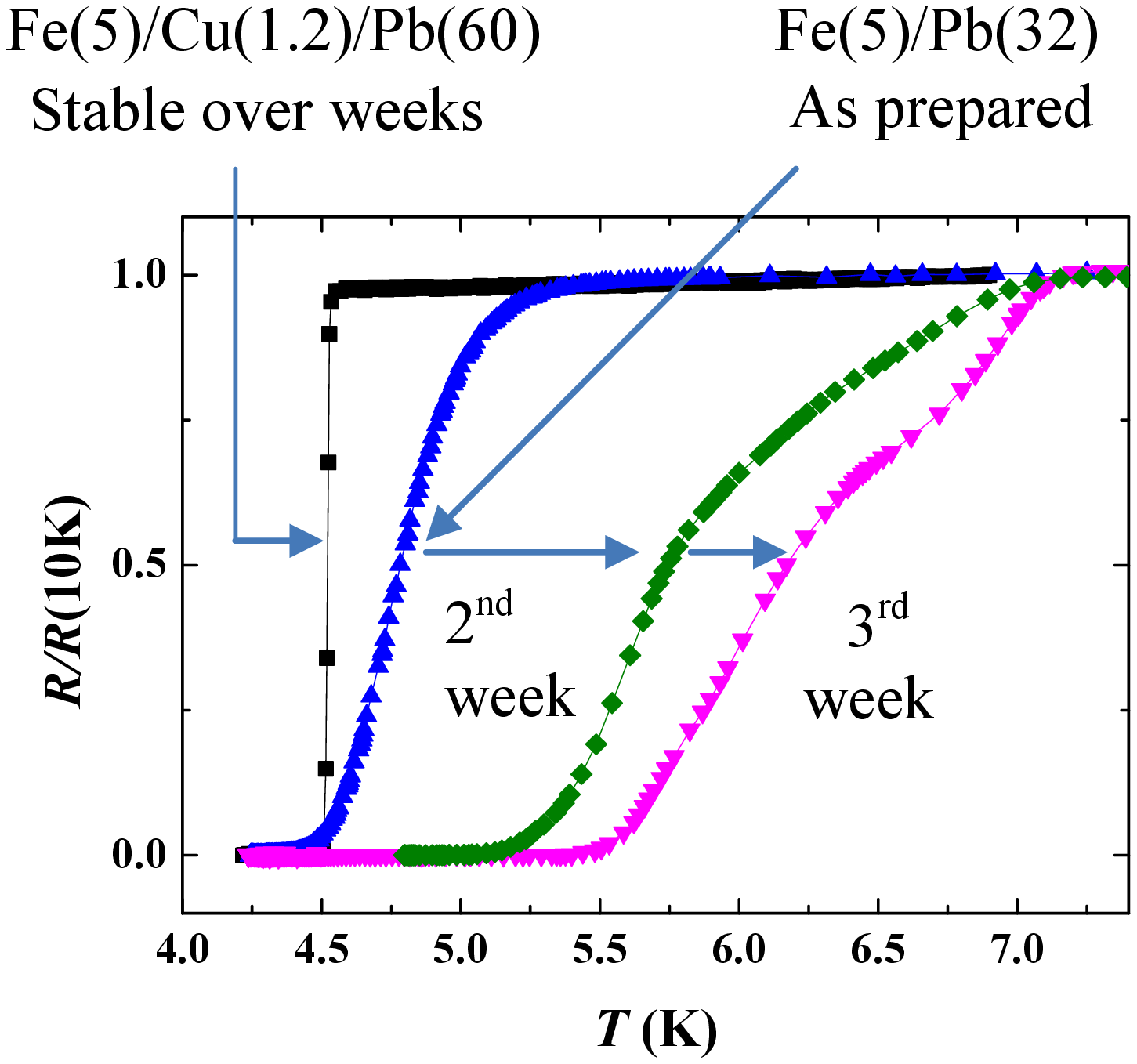}
\caption{(Color online) Time evolution of the superconducting
properties for the diffusive Fe(5 nm)/Pb(32 nm) and stable Fe(5
nm)/Cu(1.2 nm)/Pb(60 nm) structures.}
\end{figure}
structures without the AD layer the superconducting transition
width rises up to 1.5 K and the $T_c$ shifts to its bulk value of
7.2 K during the first few weeks. A light annealing at 100$^\circ$
C of the freshly prepared samples leads to their instant
degradation. Such behavior of the superconducting transition curve
can be caused by the weakening of the $T_c$ suppression through
the Fe/Pb interface. As mentioned above the most probable reason
for that is the interdiffusion of Fe and Pb atoms through the
Fe2/Pb interface. Indeed, the introduction of a thin 1.2 nm Cu
interlayer stabilizes the interface and  inhibits the broadening
with time of the superconducting transition and of the $T_c$ shift
(Fig.~2). Importantly, we find that  the Cu interlayer is almost
transparent for the Cooper pairs penetrating from S to F layer.
The penetration depth of the Cooper pairs into a nonmagnetic metal
is usually associated with the Cooper pairs coherence length
$\xi_N$ inside the N layer. In our measurements of $T_c$ of the
Cu/Pb bilayer at small Cu-layer thicknesses, the $T_c$ decreases
monotonically when $d_\mathrm{Cu}$-layer thickness increases
(Fig.~3).
\begin{figure}[h]
\includegraphics[height=0.9\columnwidth,angle=-90]{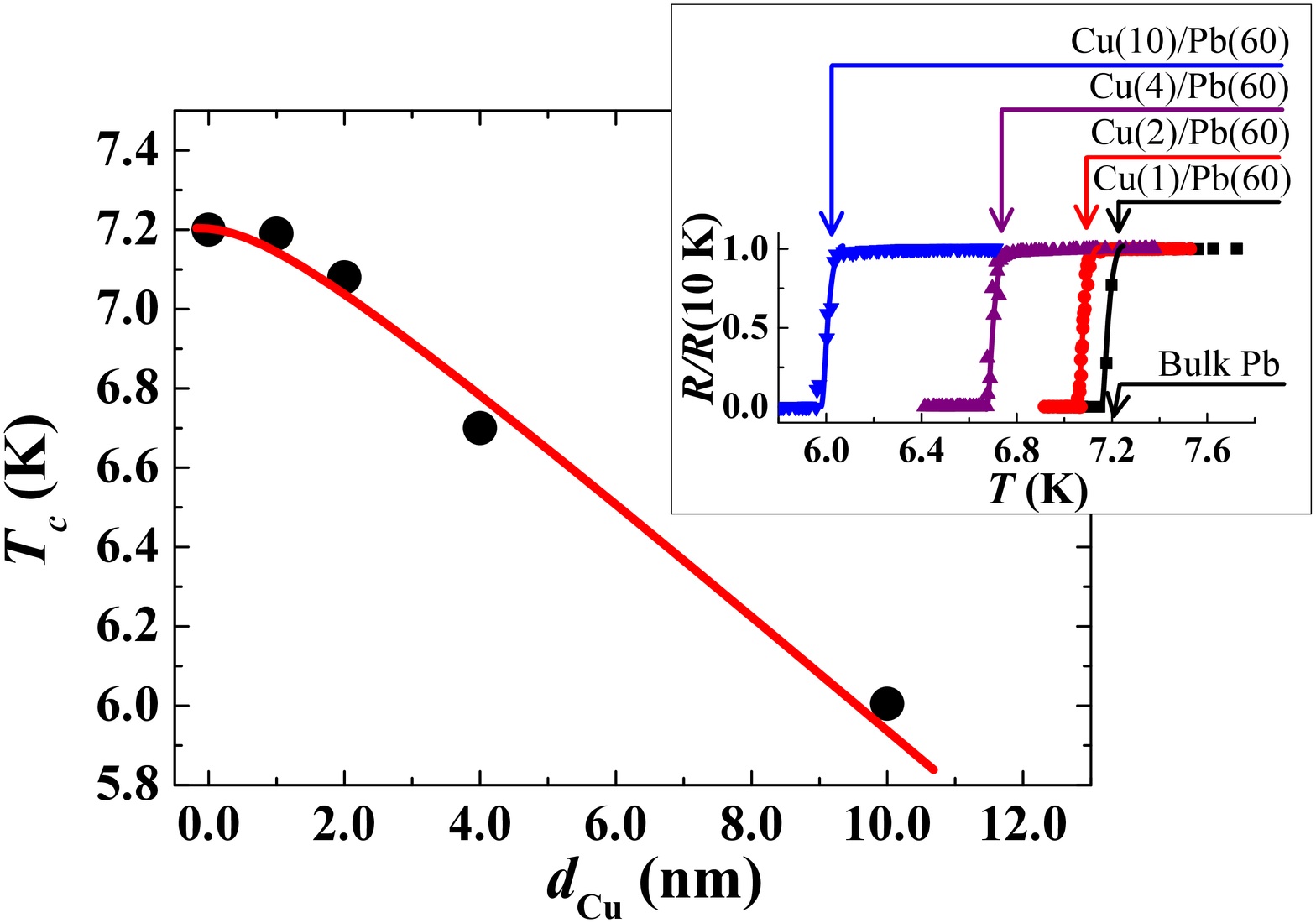}
\caption{(Color online) The $T_c$ dependence on the Cu-layer
thickness for the Cu($d_\mathrm{Cu}$)/Pb(60 nm) bilayer. Solid
line is a guide to the eye. The inset demonstrates characteristic
superconducting transition curves corresponding to the data points
on the $T_c$($d_\mathrm{Cu}$) main plot.}
\end{figure}
 This result shows that the penetration depth of the Cooper pairs  exceeds the Cu
antidiffusion layer thickness at least by 10 times. Thus one can
conclude that $d_\mathrm{Cu}=1.2$ nm is an optimal thickness of
the AD layer to stabilize the Fe/Pb interface without affecting
$T_c$.

\section{Magnetic and superconducting characterization}

\subsection{Magnetic measurements}

All samples were magnetically characterized using a standard 7T
VSM SQUID magnetometer (Fig.~4).
\begin{figure}[h]
\includegraphics[width=1.0\columnwidth,angle=-90]{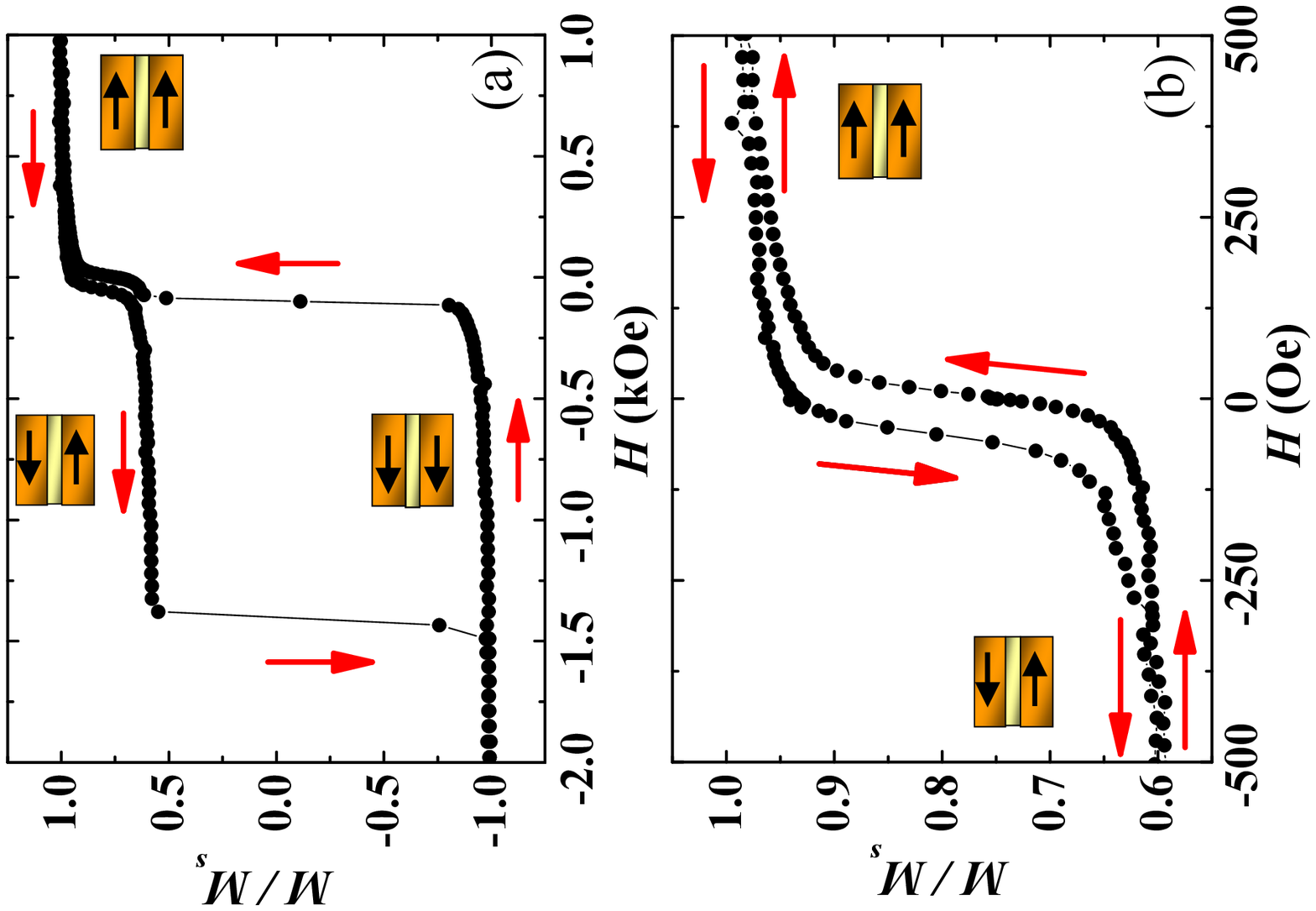}
\caption{(Color online) (a) Major hysteresis loop for the sample CoO$_x$(2.5
nm)/Fe1(3 nm)/Cu(4 nm)/Fe2(0.8 nm)/Cu(1.2 nm)/Pb(60 nm); (b) the
central part of the minor hysteresis loop for this sample due to
the reversal of the magnetization of the free Fe2 layer.}
\end{figure}
First, the sample was cooled down from 300\,K to 10\,K in the
presence of the in-plane magnetic field $+4$ kOe. At 10 K the
magnetic field was varied from $+4$ kOe down to $-6$ kOe and back
again. During this variation the in-plane magnetic moment of the
sample was measured. It turns out that for most of the samples
CoO$_x$/Fe1/Cu/Fe2/Cu/Pb the magnetic field of about $\pm 500$ Oe
is enough to get fully saturated magnetization of the free Fe2
layer while the magnetization of the Fe1 layer remains fixed up to
the operating field of the order of $-1$ kOe. This upper limit
strongly depends on the Fe1-layer thickness. If the operating
field is larger than this upper limit, then both magnetizations
start to rotate and the angle between them is not clearly defined
in the experiment. The hysteresis magnetization loop for each Fe
layer has its own saturation value, proportional to the thickness
of the Fe layer. The SQUID data were also used to correct the
Fe-layers thickness using the specific magnetic saturation value
which was of the order of the bulk value of 1730 emu/cm$^3$.

\subsection{Electrical resistivity}

The electrical resistivity measurements were performed using the
standard  four-point method. The top insulating layer was mechanically
removed at the points where the golden wires should be attached to
the surface of the sample.

The $T_c$ is defined as the midpoint of the superconducting
transition. The experimental setup for the transport measurements is
a combination of an electrical setup with a vector electromagnet
that enables a continuous in-plane rotation of the sample in
magnetic field. For a precise control of the magnetic field value a
Hall probe with a sensitivity better than $\pm 0.3$ Oe  was used. To
avoid the occurrence of the unwanted out-of-plane component of the
external field the sample plane position was always adjusted with an
accuracy better than $3^\circ$ relative to the direction of the dc
external field.

The quality of the Pb layers can be judged from the residual
resistivity ratio $RRR = R(300 K)/R(4
K)=[\rho_{ph}(300K)+\rho(4K)]/\rho (4K)$, where $R(T)$ is the
measured resistance at temperature $T$, $\rho_{ph}$(300 K) is the
phonon contribution to the specific resistivity at 300 K and
$\rho$(4 K) is the residual resistivity at 4 K. For the studied
samples $RRR$ is about 15. The high value of $RRR$ is usually
correlated with a high purity of the films. Since the room
temperature resistivity of the Pb layer is dominated by the phonon
contribution $\rho_{ph}(300K)=21$ $\mu\Omega \cdot$cm (see, e.g.,
\cite{Kittel}) one can estimate the residual specific resistivity
$\rho_0$ from the $RRR$ value. We obtain $\rho_0 = \rho$(4K)=1.47
$\mu\Omega \cdot$cm. The Pippard relations \cite{Pippard}
\begin{equation}
\sigma=e^2S_F  l /12\pi^3\hbar,\quad \gamma^e=k_B^2
S_F/12\pi \hbar  v_F,
\end{equation}
give
\begin{equation}
v_F l=\left( \frac{\pi k_B}{e}
\right)^2 \left( \frac{1}{\gamma^e \rho} \right).
\end{equation}
Here  $S_F$ is the Fermi surface area,  $l$ and $v_F$ are the mean
free path and Fermi velocity of conduction electrons, both
averaged over the Fermi surface, and $\gamma^e$ is the electronic
specific heat coefficient.  Using for Pb $\gamma^e=1.6\cdot 10^3$
erg/K$^2$cm$^3$ (Ref.~\onlinecite{Kittel}), we obtain $v_F^S l_S =
3\cdot 10^2$ cm$^2$/s and the corresponding diffusion constant in
the S layer $D_S= v_F^S l_S /3=10^2$ cm$^2$/s. In our case
$v_F^S=1.8\cdot 10^8$ cm/s \cite{Kittel}. This means that the mean
free path of conduction electrons in Pb is $l_S=17$ nm. The BCS
coherence length for Pb amounts to $\xi_0=80$ nm
\cite{Deutscher1}. The comparison of the mean free path of
conduction electrons $l_S$ with the superconducting coherence
length $\xi_0$ shows that $l_S<\xi_0$ implying the dirty limit for
the superconducting part of the system. The S-layer coherence
length for the dirty limit reads
\begin{equation}
\xi_S=\sqrt{\frac{\hbar D_S}{2\pi k_B T_{cS}}},
\end{equation}
where $T_{cS}$ is the critical temperature of the S layer. We
follow notations of Ref. \cite{Fominov3} restoring $\hbar$ and
$k_B$ in Eqs.~(3)-(4) since the are used in other formulas in the
paper. Equation~(3) gives $\xi_S=42$ nm for our samples.

We also tried to measure the residual resistivity of the Fe layers
in the thickness range corresponding to the studied spin valve
samples.  We cannot measure the partial resistivity of each layer
in the multilayer sample. Therefore, we had to measure the
resistivity of such thin layers using single layer and bilayer
films. It turns out that iron films grown directly on the MgO
substrate at room temperature become discontinuous at a thickness
below 10 nm. Therefore we prepared the set of samples MgO/Cu(4
nm)/Fe($d_\mathrm{Fe}$). In this case the quality of the iron
layer was much better, and for Cu(4 nm)/Fe(5 nm) we found that
$\rho_0^{Fe}$ of the order of 10-30 $\mu \Omega \cdot$cm. Such a
big scatter of the resistivity values is caused by different
roughnesses of the iron layers. Possibly, residual resistivity of
the Fe2 layer $\rho_0^{Fe2}$ in a real multilayer could be
different. For our analysis we shall use an ``optimistic'' value
of the residual resistivity for iron $\rho_0^{Fe}=10\ \mu \Omega
\cdot$cm. Using Eq.\ (2) we obtain $v_F^F l_{F} =10$ cm$^2$/s,
$D_F=3.3$ cm$^2$/s and $l_F=1.0$ nm. In notation of Ref.
\cite{Fominov3}, we have
\begin{equation}
\xi_F=\sqrt{\frac{\hbar D_F}{2\pi k_B T_{cS}}}.
\end{equation}
From this equation we get $\xi_F=7.5$ nm.

\section{Experimental results}
\subsection{Dependence of the superconducting transition temperature on Pb- and Fe-layers thickness}

On the first stage of experiments the magnetic part of the system
CoO$_x$/Fe1/Cu/Fe2 was optimized in order to get a well-defined
control of the mutual orientation of the Fe1 and Fe2 layers'
magnetizations. For such combination we found the optimal
thicknesses for the CoO$_x$ layer of 2.5 nm and for the Cu
interlayer of 4 nm. The pinning effect for the combination
CoO$_x$/Fe1 is not strongly affected by the CoO$_x$ thickness
variation. However artefacts in the results of magnetic
measurements may result from the incomplete magnetization pinning
when the thickness of the Co layer is not large enough to form a
uniform CoO$_x$ film on the MgO substrate. The Cu interlayer
thickness of 4 nm was found to be optimal to prevent magnetic
coupling between Fe1 and Fe2 layers at low temperatures.

The aim of the second step was the finding of the optimal Pb-layer
thickness for the observation of the S/F proximity effect. The
Pb-layer thickness should be sufficiently small to make the whole
S layer sensitive to the magnetic part of the system. Only in this
case the mutual orientation of the Fe-layers' magnetizations would
affect $T_c$ of the system. In order to define the optimal
thickness we measured the $T_c$ dependence on the Pb layer
thickness for the Fe/Pb and Fe/Cu/Pb systems (Fig.~5).
\begin{figure}[h]
\includegraphics[height=0.9\columnwidth,angle=-90]{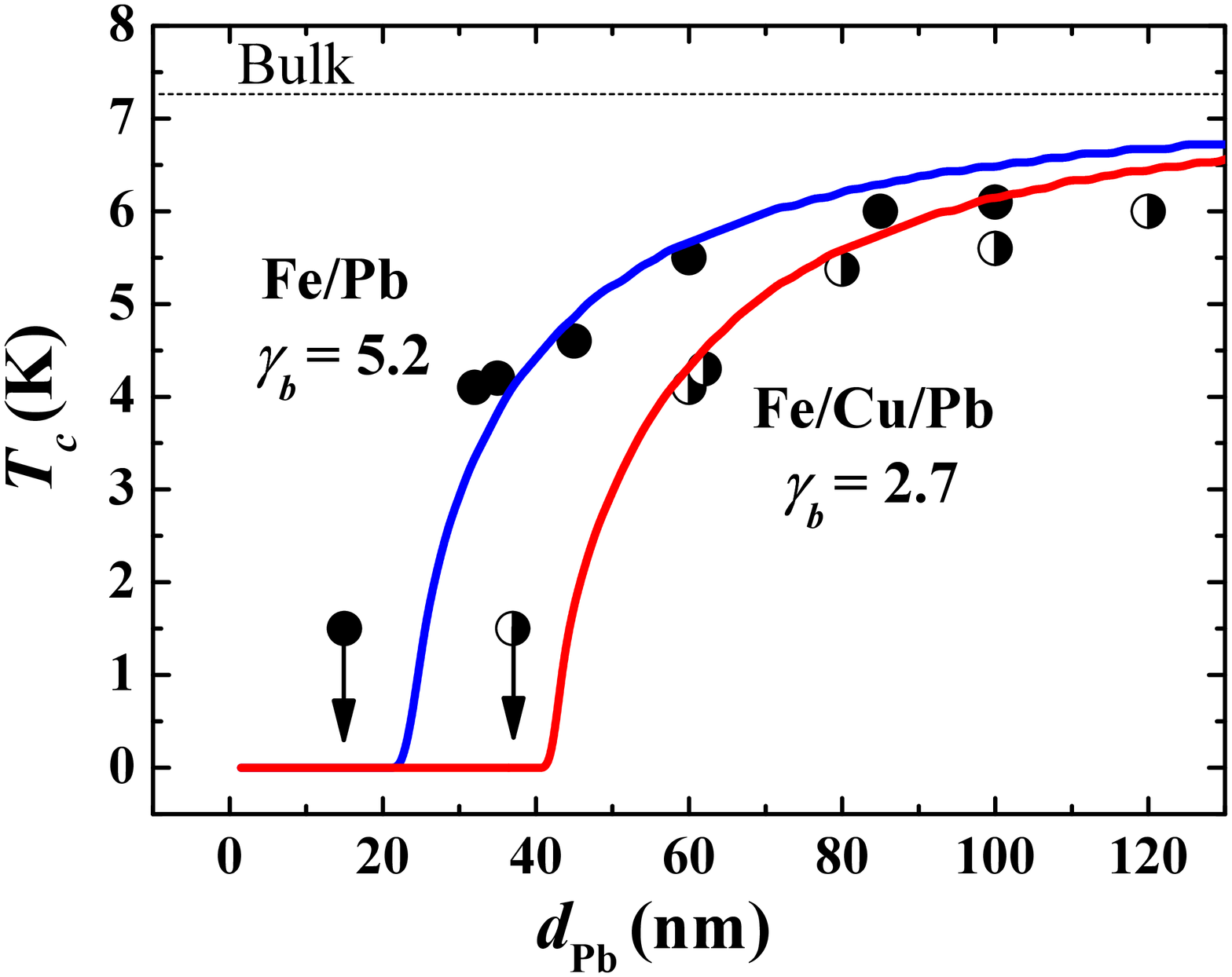}
\caption{(Color online) $T_c$ vs.\ $d_\mathrm{Pb}$ at fixed
$d_\mathrm{Fe}=5$ nm for the Fe(5 nm)/Pb bilayer (closed circles)
and Fe(5 nm)/Cu(1.2 nm)/Pb trilayer (half-opened circles). Solid
lines are theoretical fits using the model by Fominov  \textit{et
al.} \cite{Fominov3} (see Section VI).}
\end{figure}
The $T_c$ value for the Fe/Cu/Pb starts to decrease rapidly when
the thickness of the Pb layer $d_\mathrm{Pb}$ is reduced down to
80 nm. At $d_\mathrm{Pb}\leq$ 40 nm the $T_c$ is less than 1.5 K.
According to Fig.~5 the optimal thickness of the Pb layer lies
between 80 and 40 nm. At small $d_\mathrm{Pb}$ the width of the
superconducting transition curves $\delta T_c$ gets extremely
large, of the order of 0.5 K. Most likely it is caused by the
roughness of the Pb layer which become important at small
$d_\mathrm{Pb}$ where the derivative $\partial T_c/ \partial
d_\mathrm{Pb}$ is large. For the samples with $d_\mathrm{Pb}=60$
nm, we obtain $T_c = 4.5$ K with a quite small $\delta T_c $ of 50
mK. The $T_c$ for these samples is low enough in comparison with
the bulk Pb. Note that, first, the $T_c$-suppression by the 5 nm
thick iron layer for $d_\mathrm{Pb}=60$ nm is about 3 K and,
second, the thickness $d_\mathrm{Pb}=60$ nm only slightly exceeds
the superconducting coherence length $\xi_S=42$ nm. This suggests
the thickness $d_\mathrm{Pb} = 60$ nm as the optimal thickness for
our study. According to Fig.~5 for the Fe(5 nm)/Pb(60 nm) system
without Cu antidiffusion layer the $T_c$ is higher than $T_c$ for
Fe(5nm)/Cu(1.2 nm)/Pb(60 nm). For $d_\mathrm{Pb} = 60$ nm the
difference reaches 1 K. This cannot be explained by an additional
shift due to the contact Cu/Pb, since the $T_c$ difference between
Cu(1.2 nm)/Pb(60 nm) and Pb(60 nm) is less than 0.1 K (see
Fig.~3). This shift is caused most probably by a lower interface
quality in the case of Fe/Pb as compared to Fe/Cu/Pb.

\begin{figure}[h]
\includegraphics[height=0.8\columnwidth,angle=-90]{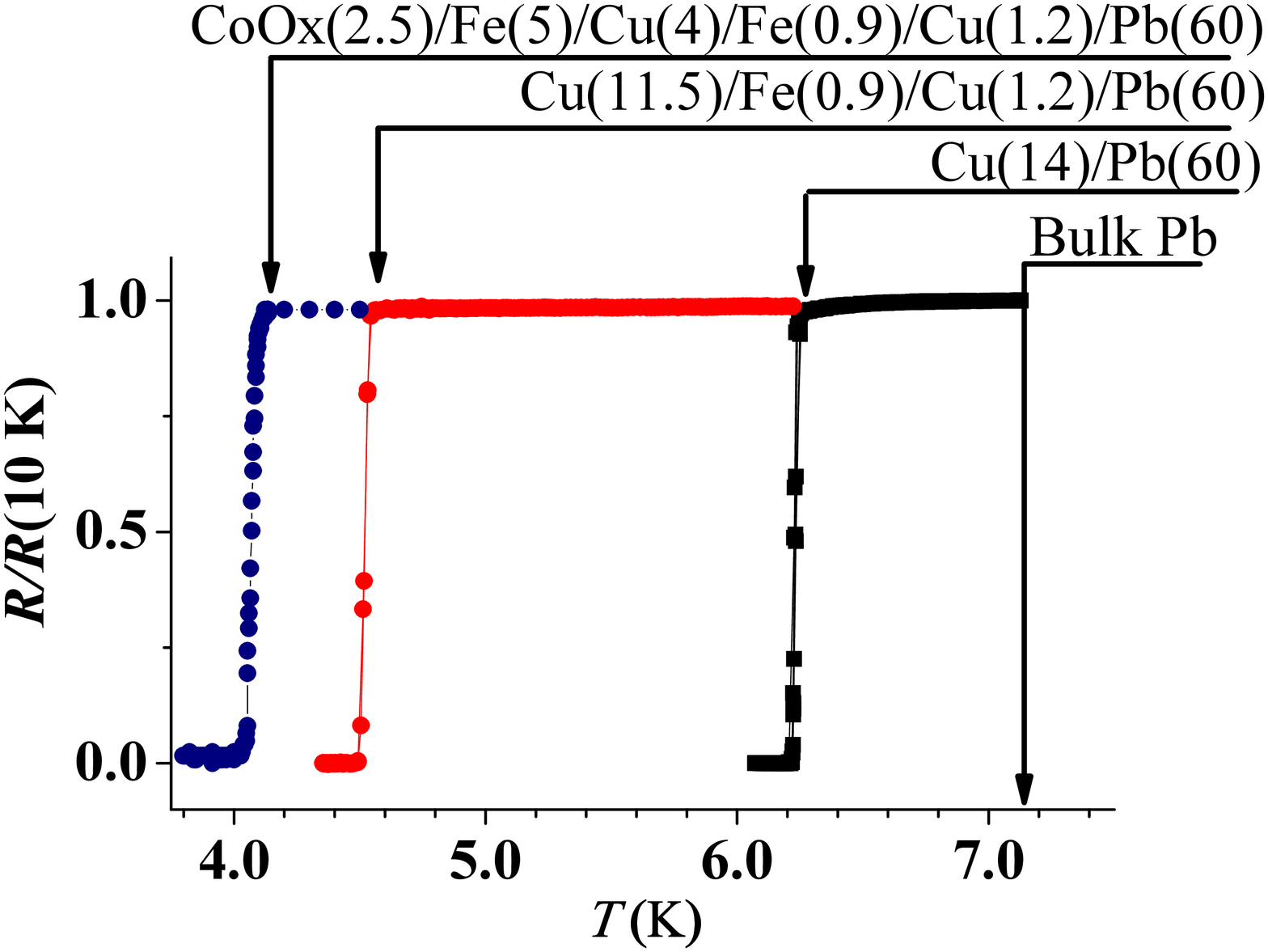}
\caption{(Color online) Evolution of the resistivity transition
curves $R(T)/R(T=4\,\mathrm{K})$ for the system
CoO$_x$/Fe/Cu/Fe/Cu/Pb due to the consecutive substitution of
magnetic layers by nonmagnetic Cu. } \label{fig5}
\end{figure}

The next logical step was to check whether the superconductivity
is influenced by both ferromagnetic layers. One of the simple ways
to check this is to study separately the influence of each
component of the layer sequence in the sample
CoO$_x$(2.5)/Fe1(4)/Cu(4)/Fe(0.5)/Cu(1.2)/Pb(60) by a stepwise
replacing of the ferromagnetic Fe layers by a non-magnetic Cu
layer with the same thickness (Fig.~\ref{fig5}).
The pair-breaking effect for the Fe layer is different than that
for Cu. This would result in a change of $T_c$. As expected, the
lowest $T_c = 4$ K was observed for the full stack of magnetic
layers CoO$_x$(2.5)/Fe1(5)/Cu(4)/Fe(0.9)/Cu(1.2)/Pb(60). When the
magnetic part CoO$_x$(2.5)/Fe1(5)/Cu(4) was replaced by the
nonmagnetic Cu(11.5) layer with the same total thickness, the
$T_c$ value increased up to 4.5 K. And finally, the substitution
of the whole magnetic part
CoO$_x$(2.5)/Fe1(5)/Cu(4)/Fe(0.9)/Cu(1.2) by Cu(14) resulted in a
significant change of $T_c$ up to 6.25 K. These results evidence
that the Cooper pairs penetrate from the superconducting Pb layer
through the Fe2 layer inside the magnetic part and sense all
layers including the Fe1 layer.

\subsection{Superconducting spin valve effect}

The magnitude of the superconducting spin valve effect $\Delta T_c =
T_c^\mathrm{AP}- T_c^\mathrm{P}$ is the measure of the difference between the
critical temperatures for AP ($T_c^\mathrm{AP}$) and  P ($T_c^\mathrm{P}$)
orientations of magnetizations of the F layers. After the field
cooling procedure at $+4$ kOe the Fe1 layer magnetization is pinned in
the positive direction by antiferromagnetic CoO$_x$ even at high
switching magnetic field up to $-1.5$ kOe (Fig.~4(a)). Also the direction
of the Fe2 layer magnetization is positive providing initial P state.
The Fe2 layer magnetization is not pinned and the application of a
small switching field in the negative direction can switch the
mutual orientation from P to AP state (Fig.~4(b)).  The saturation
field for the Fe2 film is of the order of 500 Oe. This means that
the switching field $H_0 = 500$ Oe is sufficient to sustain a
homogenous magnetization for the Fe2 layer following the switching
field direction without formation of the domain structure.

The superconducting spin valve effect $\Delta T_c(d_\mathrm{Fe2})$ was
measured for two sets of samples
CoO$_x$(2.5)/Fe1(2.5)Cu(4)Fe($d_\mathrm{Fe2}$)/Cu(1.2)/Pb(60) and
CoO$_x$(2.5)/Fe1(2.5)Cu(4)Fe($d_\mathrm{Fe2}$)/Pb(60). The first set
contains a Cu AD layer, the second one does not contain this layer.
In both sets the Fe2 layer thickness $d_\mathrm{Fe2}$ is varied while other
layers' parameters are fixed. The thickness of the Fe1 layer
$d_\mathrm{Fe1} = 2.5$ nm is much larger than $\xi_{h}$ which is less than
0.8 nm. Hence the $\Delta T_c$ is not affected by the $d_\mathrm{Fe1}$
variation. The results for these two sets with varying $d_\mathrm{Fe2}$ at
fixed $d_\mathrm{Fe1}=2.5$ nm are demonstrated in Fig.~7.

\begin{figure}[h]
\includegraphics[height=0.9\columnwidth,angle=-90]{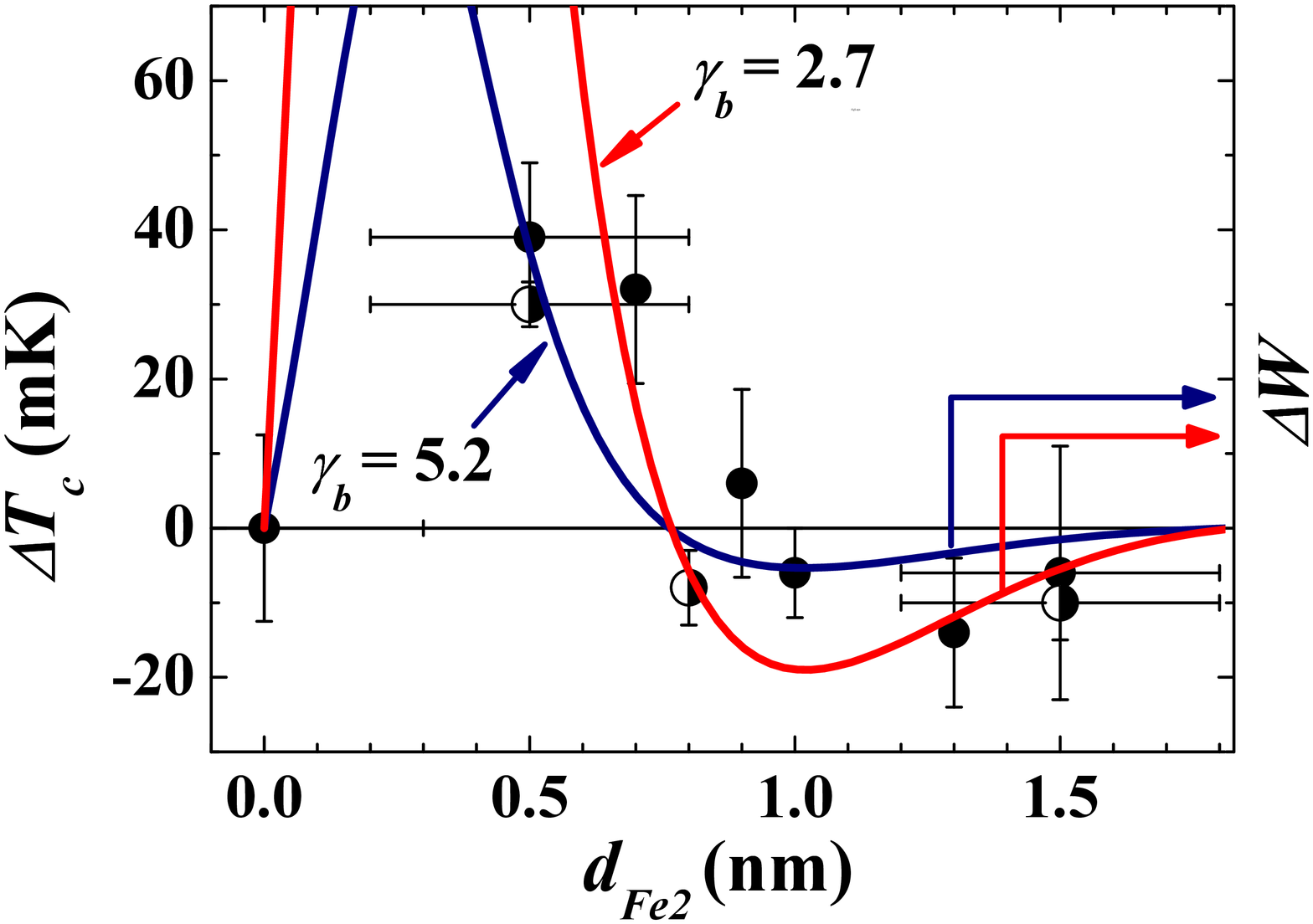}
\caption{(Color online) The dependence of the superconducting spin
valve effect magnitude $\Delta T_c$ on the Fe2-layer thickness for
CoO$_x$/Fe1/Cu/Fe2/Pb (closed circles) and for
CoO$_x$/Fe1/Cu/Fe2/Cu/Pb (half-opened circles) systems.
Theoretical calculations for $\gamma_b=2.7$ and $\gamma_b=5.2$
transparency parameter of the S/F interface are depicted by red
and blue curves (see Section VI).} \label{fig6}
\end{figure}

The $\Delta T_c(d_\mathrm{Fe2})$-dependence exhibits a
well-defined sign-changing oscillating behavior. First, at
$d_\mathrm{Fe2}=0.5$ nm the $\Delta T_c$ has a maximum of +40 mK.
The positive value implies the direct effect,
$T_c^\mathrm{AP}>T_c^\mathrm{P}$. An increase of $d_\mathrm{Fe2}$
value causes a reduction of $\Delta T_c$ down to 0 mK around
$d_\mathrm{Fe2} = 0.8 \div 1$ nm. Notably, when $d_\mathrm{Fe2}$
increases further above 1 nm, the $\Delta T_c$ falls down to a
negative minimum of -15 mK at $d_\mathrm{Fe2} \sim$ 1.3 nm and
then rises up back to 0 at $d_\mathrm{Fe2} \sim$ 1.5 nm. It is
also important to note that the oscillating behavior for this
dependence is a characteristic feature for both types of systems
containing the Fe/Pb and Fe/Cu/Pb interfaces. Similar result was
also observed in the case of CoO$_x$/Fe/Cu/Fe/In heterostructures
(Refs. \cite{Leksin2,Leksin3}).

\begin{figure}[]
\includegraphics[height=0.9\columnwidth,angle=-90]{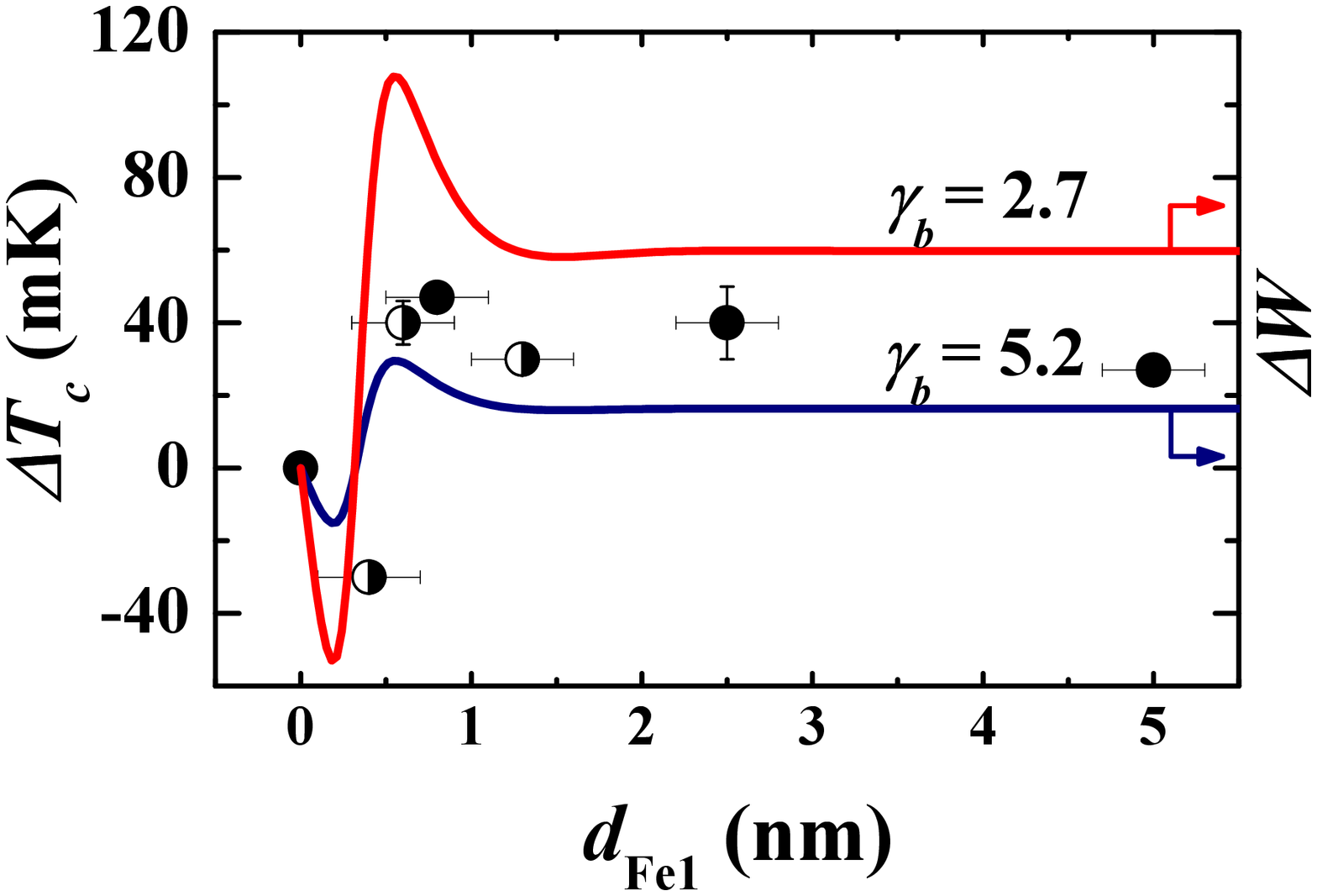}
\caption{(Color online) The dependence of magnitude of the
superconducting spin valve effect $\Delta T_c$ on the Fe1-layer
thickness. The results were obtained for Fe2 layer thickness
$d_\mathrm{Fe2} = 0.5$ nm. Closed circles and half-opened circles
correspond to CoO$_x$/Fe1/Cu/Fe2/Cu/Pb samples prepared at two
different vacuum cycles. Theoretical calculations for $\gamma_b=2.7$
and $\gamma_b=5.2$ transparency parameter of the S/F interface are
depicted by red and blue curves (see Section VI).}
 \label{fig7}
\end{figure}

In order to investigate the dependence of the magnitude of the
superconducting spin valve effect on the Fe1 layer thickness we also
studied a set of samples
CoO$_x$(2.5)/Fe1($d_\mathrm{Fe1}$)/Cu(4)/Fe($d_\mathrm{Fe2}$)/Cu(1.2)/Pb(60)
with two fixed Fe2-layer thicknesses $d_\mathrm{Fe2}=$ 0.5 and 0.9
nm. The results for the samples with $d_{Fe2}$=0.5 nm are depicted
in Fig.~\ref{fig7}.
One can see that the dependence in Fig.~\ref{fig7} have the minor
maximum of $\Delta T_c(d_\mathrm{Fe1})$ at the values of
$d_\mathrm{Fe1}$ of the order of 0.8 nm.  Qualitatively, the maximum
in $\Delta T_c(d_\mathrm{Fe1})$ is due to the optimized value of
$T_c^\mathrm{AP}$. For relatively thin Fe layers of nearly equal
thicknesses, the exchange fields in the antiparallel configuration
effectively cancel each other, leading to a maximized
$T_c^\mathrm{AP}$. When increasing $d_{Fe2}$ from 0.5 to 0.9 nm the
magnitude of $\Delta T_c$  approaches zero in accordance with
Fig.~7. With further increasing $d_{Fe2}$ the spin valve effect
change sign but still remains very small. Another interesting
feature of these dependences is the negative minimum at extremely
small Fe1 layer thickness (Fig.~8(a)). This kind of oscillations of
$T_c(d_\mathrm{Fe1})$ is also due to the interference effects as it
was in the case of the $T_c(d_\mathrm{Fe2})$ dependence (Ref.
\cite{Leksin2}). For our sample growth technique, the thinnest
uniform Fe layers which can be prepared has, in the best case, a
thickness of 0.5 nm and not less. However, for the CoO$_x$/Fe1 part
we observed a partial oxidation of the Fe layer. The oxidation depth
of the Fe1 layer is about 0.4-0.6 nm. Thus, preparation of a sample
stack CoO$_x$/Fe1(1)/Cu/Fe2/Pb actually yields the stack with the
effectively smaller thnickness of the Fe1 layer of 0.4 - 0.6 nm
CoO$_x$/Fe1(0.4-0.6)/Cu/Fe2/Pb. We also successfully used this
effect to reduce the thickness of the Fe1 layer down to 0.3-0.4 nm
(Fig.~8(a)).

\begin{figure}[]
\includegraphics[height=0.7\columnwidth,angle=-90]{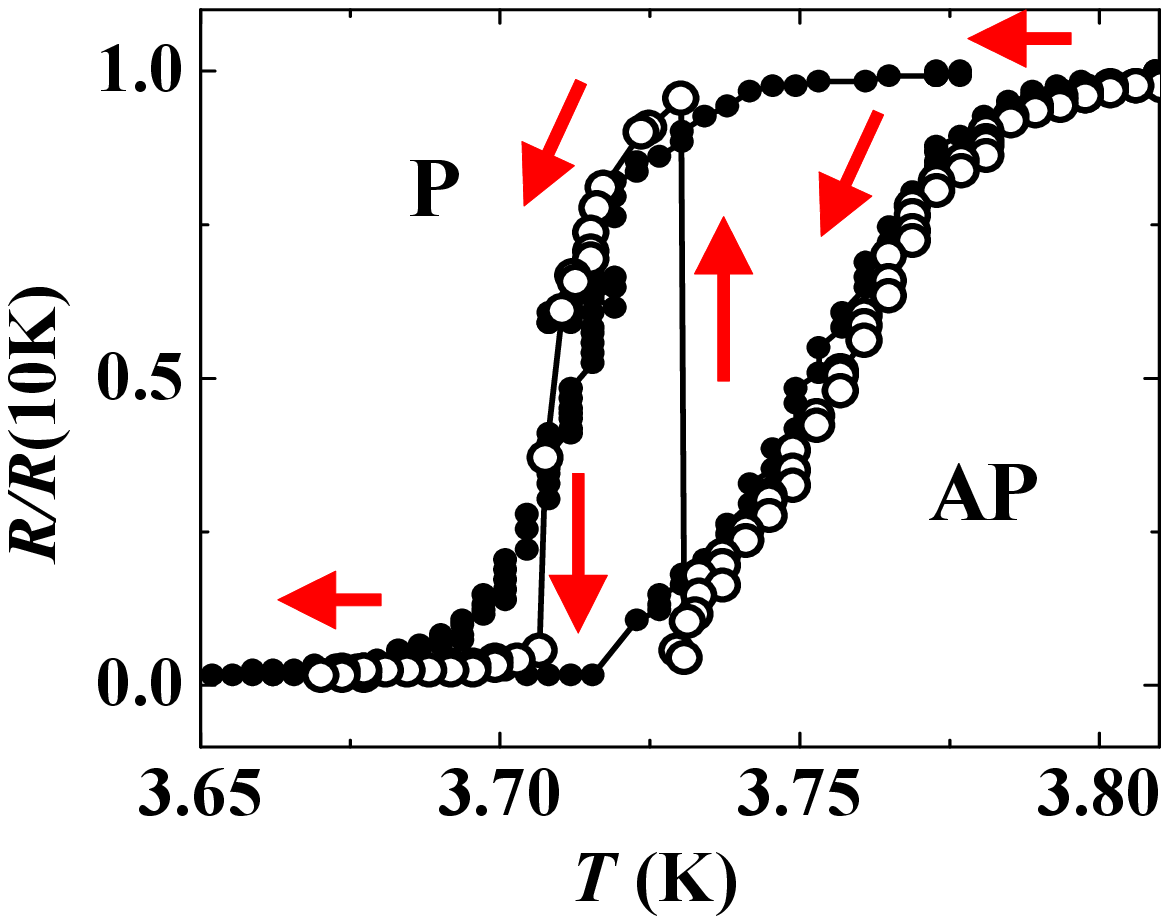}
\caption{(Color online) ($\bullet$) The superconducting transition
curves for CoO$_x$/Fe1(0.8)/Cu(4)/Fe(0.5)/Cu(1.2)Pb(60) for P and
AP mutual orientation of the Fe1 and Fe2 layers' magnetizations.
($\circ$) Instantaneous switching between the superconducting and
normal states for this sample by switching between AP and P
orientations.}
 \label{fig8}
\end{figure}

The full switching of the supercurrent was achieved for the sample
CoO$_x$/Fe1(0.8)/Cu(4)/Fe2(0.5)/Cu(1.2)/Pb(60) which has the maximum
$\Delta T_c$ according to Fig.~\ref{fig7}(a). The difference in
$T_c$ between AP and P states for this sample is clearly seen in
Fig.~\ref{fig8}.
The superconducting transition temperature  for the AP orientation
of the magnetizations occurs at a temperature exceeding $T_c$ for
the P orientation by 40 mK, which is of the order of the
superconducting transition width $\delta T_c$. This opens a
possibility to switch off and on the superconducting current
flowing through our sample almost completely within the
temperature range corresponding to the $T_c$-shift by changing the
mutual orientation of the magnetizations of Fe1 and Fe2 layers. To
demonstrate this we have performed  resistivity measurements of
the sample by sweeping slowly the temperature within the $\Delta
T_c$ and switching the magnetic field between $+1$ and $-1$ kOe.
We note that the full switching of the supercurrent using the
proximity effect was observed for the first time in the
CoO$_x$/Fe/Cu/Fe/In system \cite{Leksin1}. However, using Pb
instead of In allowed us to improve the critical parameters of the
system. In particular, we have increased the operating temperature
from 1.4 K for Fe/In-based systems up to 3.7 K for Fe/Cu/Pb based
systems which is near the $^4$He boiling point.

\subsection{Triplet spin valve effect}

The triplet spin valve effect in F1/F2/S systems manifests itself in
a non-monotonic variation of the $T_c$ value upon a continuous
change of the angle $\alpha$ between magnetizations of the F1 and F2
layers from 0 to 180$^\circ$, with a minimum of $T_c$ corresponding
to a noncollinear orientation \cite{Fominov2}. According to theory,
the minimum is most pronounced if it takes place near the orthogonal
orientation of the F-layers' magnetization. This feature is a
fingerprint of the LRTC arising in the system. This component is
generated from the short-range triplet component at the F1/F2
interface. In its turn, the short-range triplet component with zero
projection of spins is generated from the conventional singlet
Cooper pairs penetrating from the S layer into the magnetic part of
the structure. The result of this proximity effect is a decrease of
$T_c$ due to ``leakage'' of Cooper pairs from the S layer. The
``leakage channel'' caused by the generation of LRTC at the F1/F2
interface should be sensitive to the number of Cooper pairs reaching
the interface. Hence, the magnitude of the triplet spin valve effect
is sensitive to the F2 layer thickness. The penetration depth of the
short-range components inside the F2 layer is of the order of
$\xi_h$ and in the case of Fe it is extremely small, $\xi_h \lesssim
1$\,nm. Increasing thickness of the F2 leads to a smaller amplitude
of the short-range triplet components at the F1/F2 interface and
hence to the reduction of the amplitude of the generated LRTC.

Our previous experiments have shown that Fe/Pb system is very
promising for the observation of the triplet spin valve effect
\cite{Leksin4}. However, the stability of such samples was
limited. This problem is solved in the present work by introducing
an AD layer between the Fe and Pb layers. This also results in an
improvement of the superconducting parameters of the system such
as the width of the superconducting transition which decreased
from 1 K down to 50 mK. It is therefore important to investigate
the triplet spin valve effect in the systems with a stable
interface and to compare these results with previously studied
Fe/Pb systems \cite{Leksin4}. The sample composition was chosen as
CoO$_x$(2.5)/Fe(3)/Cu(4)/Fe2($d_\mathrm{Fe2}$)/Cu(1.2)/Pb(60).
Below we focus on one of the representative samples,
CoO$_x$(2.5)/Fe(3)/Cu(4)/Fe2(0.8)/Cu(1.2)/Pb(60). The angular
dependences of the difference between the critical temperature
$T_c(\alpha)$ at the angle $\alpha$ and at the parallel
configuration of magnetizations $\alpha=0$ ($T_c(0)$) for this
sample are shown in Fig.~10.
\begin{figure}[]
\includegraphics[height=0.7\columnwidth,angle=-90]{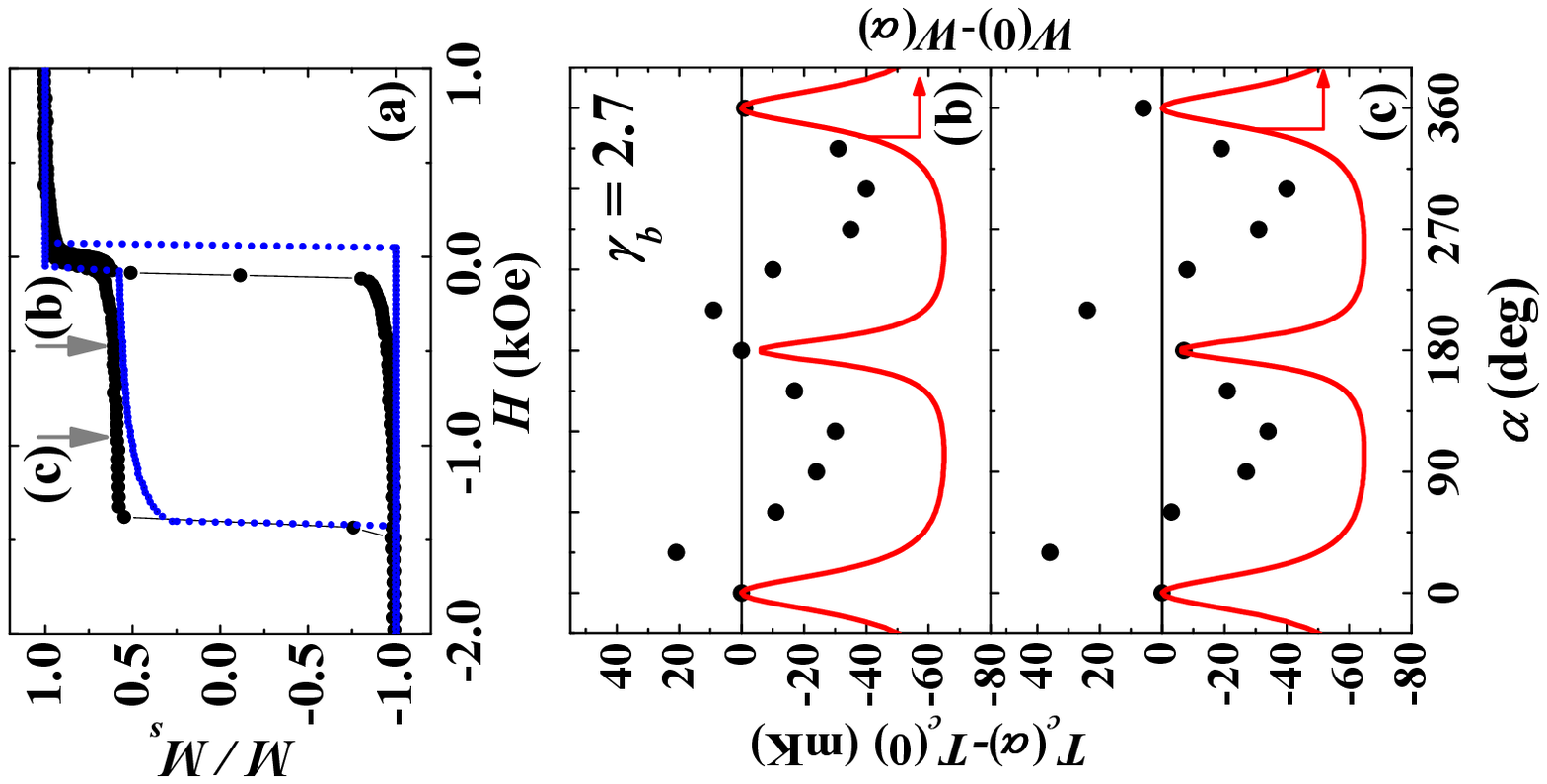}
\caption{(Color online) (a) Hysteresis loop for the sample
CoO$_x$(2.5)Fe1(3)/Cu(4)/Fe2(0.8)/Cu(1.2)/Pb(60). Arrows on (a)
show the rotating field values used when measuring the angular
dependences shown on (b) and (c). The $T_c (\alpha)-T_c(0)$
dependence for the rotating field values of 500 Oe (b) and 1 kOe
(c). Theoretical fit of the hysteresis loop is given by dotted
blue curve. Calculations of $W(0)-W(\alpha)$ using the triplet
spin valve model (b) and (c) are presented by a solid red curve
(see Section VI). }
\end{figure}
To compare the magnetic and superconducting properties of the
sample, Fig.~10 also shows the magnetic hysteresis loop measured
at 10 K and the angular dependences of $T_c$ at the magnetic field
value $H=500$ Oe and 1 kOe. After the in-plane 4 kOe magnetic
field cooling procedure the magnetizations of the Fe1 and Fe2
layers are aligned parallel $\alpha = 0$. The Fe1 layer is fixed
in a certain direction. The magnetization of the Fe2 layer is free
and can be rotated by a small magnetic field starting from its
saturation field value of $200 \div 250$ Oe. To be completely sure
that we exceed the saturation field we used the field strengths
$H=500$ Oe (b) and 1000 Oe (c) for the rotation of the Fe2 layer
magnetization. Certainly, the assumption that the magnetization of
the Fe1 layer does not deviate from its initial position during
the magnetic field rotation has to be checked more carefully. A
slight deviation of the Fe1 layer magnetization from its initial
field-cooling orientation is possible even at 500 Oe rotating
field. In addition the maxima in the angular dependence occur at
the angles which do not coincide with collinear configuration of
magnetizations. In Section VI we introduce corrections to angle
$\alpha$. In particular, we consider the anisotropy produced in
the Fe1 layer by a contact with antiferromagnetic CoO$_x$ layer as
well as the deviation of the initial direction of magnetization of
the Fe1 layer from the direction of the cooling field.

Figures~10(b) and~10(c) show a clear non-monotonic dependence of
$T_c$ on the angle $\alpha$. The minimum in $T_c$ occurs at
noncollinear configuration of magnetizations. In order to
characterize the magnitude of the triplet spin valve effect let us
introduce the value $\delta T_c(\pi/2) = T_c(\pi/2) -
[T_c(0)+T_c(\pi)]/2$. Fig.~11 shows the dependence of this value on
the Fe2-layer thickness. One can see from this figure that the
results for
\begin{figure}[h]
\includegraphics[height=0.8\columnwidth,angle=-90]{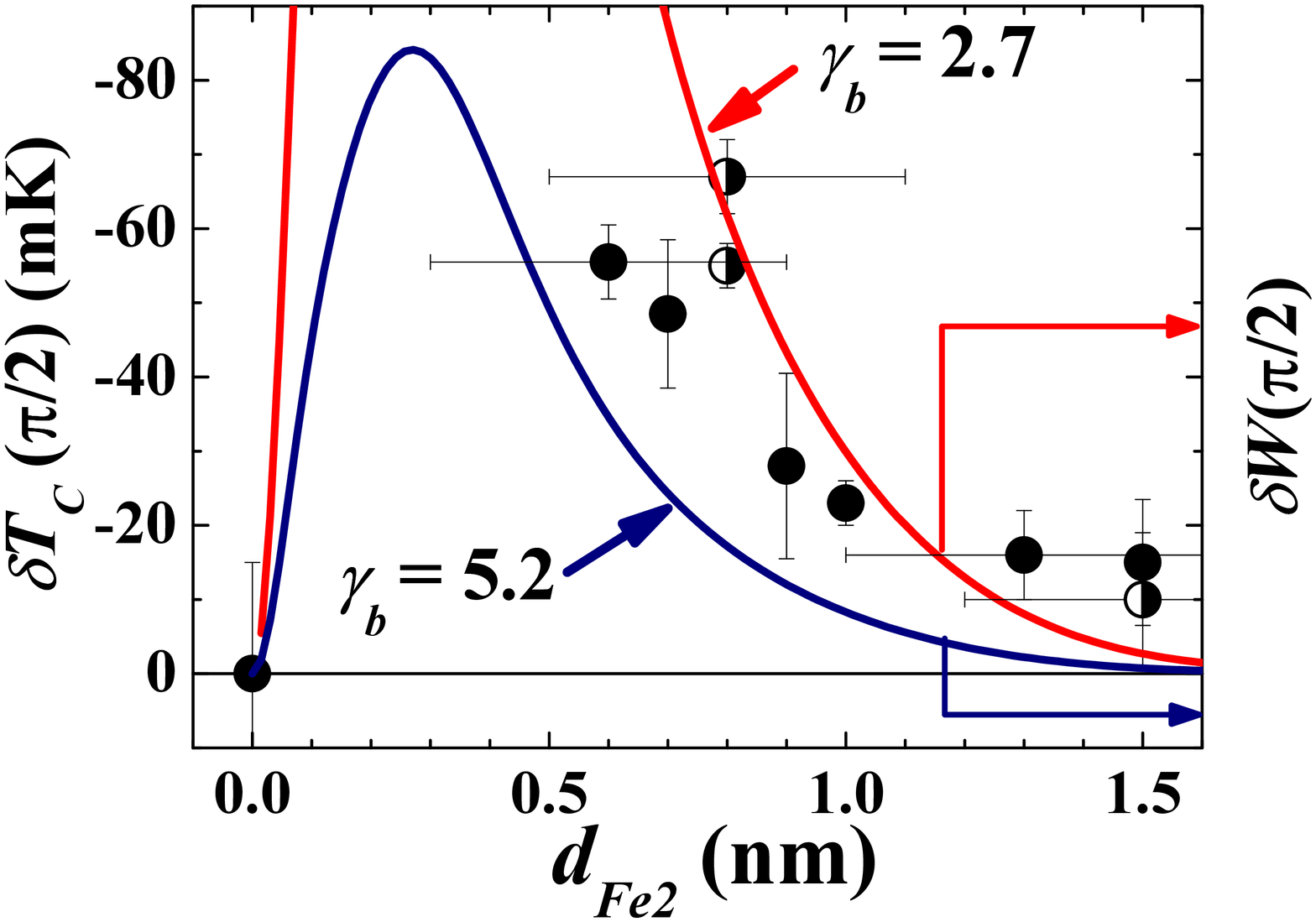}
\caption{(Color online) Dependence of the characteristic value of
the triplet spin valve effect, $\delta T_c(\pi/2)$, on the
Fe2-layer thickness for
CoO$_x$(2.5)/Fe1(2.5)/Cu(4)/Fe2($d_\mathrm{Fe2}$)/Cu(1.2)/Pb(60)
(half-opened circles) and
CoO$_x$(2.5)/Fe1(2.5)/Cu(4)/Fe2($d_\mathrm{Fe2}$)/Pb(60) (closed
circles). Theoretical calculations of $\delta W (\pi/2)$ for
$\gamma_b=2.7$ and $\gamma_b=5.2$ transparency parameter of the
S/F interface are depicted by red and blue solid curves (see
Section VI).}
\end{figure}
the set of samples based on the stabilized interface Fe/Cu/Pb
(half-opened circles) reproduce the results for Fe/Pb (closed
circles). The $\delta T_c(\pi/2)$ rises up to 67 mK when the
$d_\mathrm{Fe2}$ decrease down to 0.8 nm. An increase of
$d_\mathrm{Fe2}$ leads to a decrease of the effect. The physics of
this decrease is obviously related to the suppression of the singlet
components' amplitude at the Fe1/Fe2 interface which serves as a
source for the triplet components. We recall that the characteristic
depth of the decay of the singlet component function in the F layer
is small, $\xi _h \sim 0.8$ nm. Therefore the strongest effect is
expected for the smallest thickness $d_\mathrm{Fe2}$.

\section{Theory and discussion}

Let us start from the theoretical model \cite{Fominov2} which we
use for the analysis of our experimental results. The authors of
this model have studied the critical temperature of an F/F/S
trilayer at an  arbitrary angle between the in-plane
magnetizations, which makes it necessary to take the LRTC into
account. They reduce the problem to the form which allows to apply
a general numerical method developed in Ref. \cite{Fominov3}. The
dirty limit was considered which was described by the Usadel
equations. The linearized Usadel equations contain only the
anomalous Green function. The boundary conditions for this
function at the S/F interface contain the material matching
parameter $\gamma$ and the transparency parameter $\gamma_b$
defined as: 
\begin{equation}
\gamma=\frac{\rho_S \xi_S}{\rho_F \xi_F},\quad  \gamma_b=\frac{R_b
\mathcal A}{\rho_F \xi_F}.
\end{equation}
Here $\rho_S$ and $\rho_F$ are the normal-state resistivities of
the S and F layers; $R_b$ is the normal-state resistance of the
S/F boundary; $\mathcal A$ is its area;  the coherence length in
the S and F layers $\xi_S$ and $\xi_F$ are determined by Eqs.\ (3)
and (4). The value of $\gamma_b = 0$ corresponds to the fully
transparent S/F interface.

From the measured values of the residual resistivity of S and F
layers and the values of $v_F l$ (Eq.~(2)) estimated from
Pippard relations (Eq.~(1)) we calculate the parameters of the
studied systems presented in rows 1-13 of the Table~1.
\begin{table}
\caption{Parameters used for the fitting of the theory to the
experimental results}
\begin{center}
\begin{tabular}{|c|c|c|c|} \hline
 No. &Parameter &\ \ Fe/Pb\ \  &\ Fe/Cu/Pb\   \\ \hline
1 & $\rho_S$, $\mu \Omega \cdot$cm &1.47 &1.47  \\
\hline
2 &$v_F^S$, cm/s  &1.8$\cdot 10^8$  & 1.8$\cdot 10^8$   \\
\hline
3 &$v_F^S l_S$, cm$^2$/s & 300 & 300  \\
\hline
4 & $D_S$, cm$^2$/s &100 &100  \\
\hline 5& $\xi_S$, nm & 42&42\\
\hline
6&$l_S$, nm& 17&17\\
\hline
7&$\rho_F$, $\mu \Omega \cdot$cm &10 &10\\
\hline
8& $v_F^F$, cm/s&10$^8$ &10$^8$\\
\hline
 9&$ v_F^F l_F $, cm$^2$/s&10&10\\
 \hline
 10&$D_F$, cm$^2$/s&3.3&3.3\\
 \hline
 11&$\xi_F$, nm&7.5&7.5\\
 \hline
 12&$l_F$, nm&1.0&1.0\\
 \hline
 13&$\gamma$&0.78&0.78\\
 \hline
 14&$\gamma_b$&5.2&2.7\\
\hline
\end{tabular}
\label{tc_tab}
\end{center}
\end{table}
We start our theoretical analysis from the $T_c(d_\mathrm{Pb})$
dependences at the fixed value of $d_\mathrm{Fe}=5$ nm shown in
Fig.~5. At large Pb-layer thickness, $T_c$ slowly decreases with
decreasing $d_\mathrm{Pb}$. Below $d_\mathrm{Pb}=60$ and 35 nm for
Fe/Cu/Pb and Fe/Pb, correspondingly, the $T_c$ decreases abruptly;
at $d_\mathrm{Pb}<38$ nm and $d_\mathrm{Pb}<15$ nm, respectively,
superconductivity vanishes (1.5 K is the minimum temperature in
our experimental setup).

The theory assumes the dirty-limit conditions for both F and S
layers: $l_{F,S} \ll \xi_{F,S}$. The critical thickness of the S
layer $d_S^\mathrm{crit}$ is defined as the thickness below which
there is no superconductivity in the S/F bilayer:
$T_c(d_S^\mathrm{crit})=0$. In accordance with the theory
\cite{Fominov3} the explicit result for $d_S^\mathrm{crit}$ can be
obtained in the limit $(\gamma/\gamma_b)(d_S/\xi_S)\ll 1$ as
\begin{equation}
\frac{d_S^\mathrm{crit}}{\xi_S} = 2\gamma_E \frac{\gamma}{\gamma_b},
\end{equation}
where $\gamma_E \approx 1.78$ is the Euler constant. Fig.~5 shows
that for the Fe/Pb bilayers $d^\mathrm{crit}_S=22$ nm and for the
Fe/Cu/Pb system $d^\mathrm{crit}_S=42$ nm.   Thus, for the Fe/Pb
bilayer we obtain $d^\mathrm{crit}_S/\xi_S\simeq 0.52$ and for
Fe/Cu/Pb $d^\mathrm{crit}_S/\xi_S\simeq 1$. In accordance with
Eqs.~(5) and (6) we get for Fe/Pb $\gamma_b=5.2$ and for Fe/Cu/Pb
$\gamma_b=2.7$. These values are presented in row 14 of Table~1.
Using these $\gamma_b$ values we obtain theoretical fits shown in
Fig.~5 in accordance to the theory \cite{Fominov3}.  Thus all
parameters of the studied samples are already determined from our
transport measurements together with the data on the critical
thickness of the S layer $d_S^\mathrm{crit}$. At the same time, the
dirty-limit conditions also require $l_F < \xi_h$ and $l_F < d_F$,
while all these length scales turn out to be of the same order for
our samples. Therefore, the dirty-limit proximity theories
\cite{Fominov3,Fominov2} are at the border of their applicability
and while we expect them to capture basic qualitative features of
our experimental results, the adequate quantitative analysis of our
measurements in the framework of those theories is not guaranteed.
So, the purpose of our theoretical analysis is to describe general
qualitative tendencies that the experiment demonstrates.

Now we analyze consistently our experimental results on the full
structures. Indeed, as it was supposed in Section V A, the shift
of $T_c(d_\mathrm{Pb})$ dependence for the Fe/Pb system in
comparison with the Fe/Cu/Pb system is caused by the lower
interface quality in the former case. The pair-breaking effect of
the magnetic part and hence the $T_c$ strongly depend on the S/F
interface quality. The effect is strongest for the interface of a
good quality since Cooper pairs can easily penetrate from the S to
the F part of the structure, where they experience the
pair-breaking effect of the exchange field. A good S/F interface
quality should therefore generally lead to the maximized effect of
the $T_c$ suppression. In the opposite limit of an almost
impenetrable interface, the Cooper pairs remain in the S layer and
the pair-breaking effect is absent, hence $T_c$ is expected to be
maximized.

The quality of the interface depends on several parameters. The most
obvious one is the interface transparency for the electrons. The
mutual interaction of two contacting layers depends also on the matching
of conductivities (the condition for the derivatives of the Green
function in the Kupriyanov-Lukichev boundary conditions \cite{KL})
and matching of the Fermi surfaces. Mismatches of this sort
influence the transmission of electrons through the interface.

Let us start the numerical analysis of our results. In Ref.
\cite{Fominov2}, the anomalous Green function $\hat F$ was
expanded in the basis of the Pauli matrices $\hat \sigma_i$, $i=1,
2, 3$ and the unity matrix $\hat \sigma_0$. It can be shown that
in the case of in-plane magnetizations, the solution has the form
\begin{equation}
\hat F=f_0\hat \sigma_0+f_2\hat \sigma_2+f_3\hat \sigma_3.
\end{equation}
The $f_0$ component is real, while $f_2$ and $f_3$ are imaginary.
The order parameter $\Delta(x)$ only enters the equation for the
$f_0$ component. Calculations show that the effective boundary
condition for $f_0$ reads
\begin{equation}
\xi_S \frac{d f_0}{dx} = W f_0.
\end{equation}
All information about the F layers is contained in a single
function $W$. Knowledge of $W$ is already sufficient to draw
general conditions about the behavior of $T_c$. Qualitatively it
can be concluded that the larger $W$, the stronger $T_c$ is
suppressed. Thus, qualitative features of F1/F2/S spin valve
system can be understood from calculation of $W$. It is necessary
to note that the real fitting is done in Fig.~5 only. In other
cases we do not even plot $T_c$ but only $W$, which does not take
into account the parameters of the superconducting layer.

First, we analyze theoretically the experimental data shown in
Fig.~5. The estimated values of $\gamma_b$ we used as a starting
point in optimization of the fitting the theory to the
experimental results. One can see from Fig.~5 that the theory fits
experimental data well. As it was expected, the intermediate Cu
layer improves the interface transparency parameter from
$\gamma_b=5.2$ up to $\gamma_b=2.7$ \cite{Note1}.

The second step was to apply these microscopic parameters for a
theoretical analysis. Our theoretical model is based on the triplet
spin valve model by Fominov \textit{et al.} \cite{Fominov2} and is
extended to the case of finite S/F transparency.  Thus, in Figs.~8
and~7 the experimental dependence of $\Delta T_c$ on
$d_\mathrm{Fe1}$ and $d_\mathrm{Fe2}$ can be described by the
dependence of $\Delta W = W(0) - W (\pi)$ on $d_{F1}$ and $d_{F2}$,
respectively. At the same time the angular dependence
$T_c(\alpha)-T_c(0)$ depicted in Fig.~10 is expected to correlate
with  $W(0)-W(\alpha)$. The amplitude of the triplet spin valve
effect in Fig.~11 is expected to correlate with $\delta W(\pi
/2)=[W(0)+W(\pi)]/2-W(\pi/2)$.

The results in Fig.~10 were obtained using operating magnetic
fields of 0.5 kOe (b) and 1 kOe (c). This means that at the
rotating field of 1 kOe the magnetization of the fixed Fe1 layer
may slightly change its direction comparing with the initial one.
In this case the true angle $\alpha$ between $M_{Fe1}$ and
$M_{Fe2}$ differs from the rotation angle of the magnetic field.

We took this effect into account by applying a simple model which
assumes an AF/F1/F2 sandwich with two uncoupled ferromagnetic
films F1 and F2 in a rotating magnetic field. The magnetization of
the F2 film is completely free and follows the rotating field
direction. The F1 film has uniaxial anisotropy produced by a
contact with antiferromagnetic film AF (see \cite{Meiklejohn}).
The energy function for the F1 layer reads as follows:
\begin{equation}
E = K_u \sin^2\theta - K_{ud}\cos \theta - MH
\cos(\varphi-\theta).
\end{equation}
Here $M$ is its magnetic moment, $H$ is the external rotating
magnetic field, $\varphi$ is the angle between $H$ and the easy axis
produced by the antiferromagnetic layer, $\theta$ is the angle
between $M$ and the easy axis. $K_u$ and $K_{ud}$ are the uniaxial
and unidirectional anisotropy constants, respectively. It is
necessary to note that $K_u$ is much larger than that for a single
Fe layer because of the proximity to the  antiferromagnet. In its
turn, $K_{ud}$ is determined by the exchange bias caused by the
antiferromagnet. Minimizing $E$ at a given direction $\varphi$ of
the external field, we find deviation $\theta$ of the magnetization
from the easy axis. The true angle between the magnetizations of the
Fe1 and Fe2 layers is then given by $\alpha=\varphi-\theta$. The
best simulation of the hysteresis loop in Fig.~10 (a) is depicted by
the dotted blue curve. The result of this simulation was used  to
determine the real $\alpha$ value for the $T_c(\alpha)-T_c(0)$
angular dependences. This effect leads to distortion of the angular
dependence of $T_c(\alpha)-T_c(0)$ but does not shift $\alpha=0$ and
$\pi$ positions.

However, experimentally the $T_c$ maxima have offsets from these
positions. This implies a possible existence of the real difference
between the direction of the cooling field and the direction of the
magnetization of the Fe1 layer. To check such a possibility we have
performed ferromagnetic resonance (FMR) measurements of the sample
CoO$_x$/Fe1/Cu/Fe2/Cu using a standard X-band (10\,GHz) electron
spin resonance spectrometer. The FMR data at room temperature show
that the in-plane easy axis which does not coincide with the long
side of the rectangular shaped sample does exist already. Upon
cooling the sample in an external field directed along the long side
of the rectangular shaped samples (as we for the data shown in
Fig.~10) the magnetization of the Fe1 layer deviates from the
direction of the cooling field by 5-10 degrees. We suppose that the
easy axis of the magnetization is induced by residual magnetic field
in our vacuum system during the preparation process. Actually, the
shift of the $\alpha$ angle by 8 degrees and corrections of this
angle according to Eq.~(9) leads to a certain improvement of the
agreement between theory and experiment. The dependence
$T_c(\alpha^{cor})-T_c(\alpha^{cor} =0)$ on the corrected angle
$\alpha^{cor}$ is shown in Figs.~12a and 12b. Now the maxima of the
experimental angular dependence coincide with P ($\alpha^{cor}=0$)
and AP ($\alpha^{cor}=\pi$) configuration of magnetizations.
\begin{figure}[h]
\includegraphics[height=0.8\columnwidth,angle=-90]{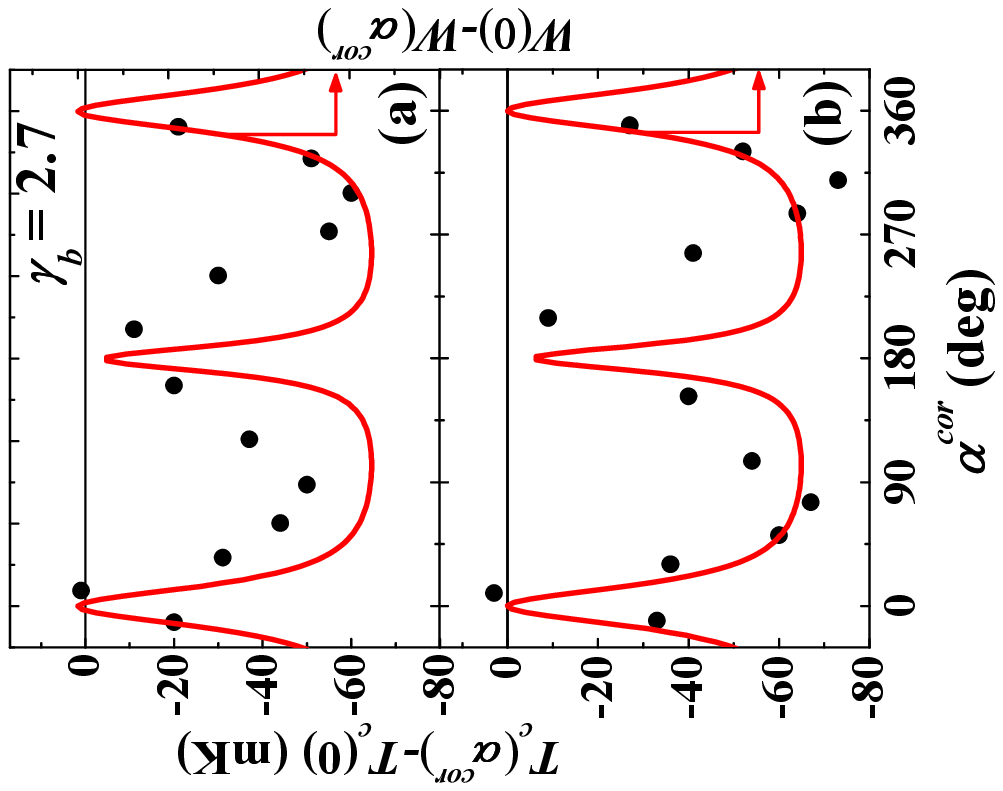}
\caption{(Color online) The $T_c (\alpha^{cor})-T_c(0)$ dependence
for the rotating field values of 500 Oe (a) and 1 kOe (b).
Calculations of $W(0)-W(\alpha^{cor})$ using the triplet spin
valve model (a) and (b) are presented by solid red curves.}
\end{figure}

The analysis of the fitting results demonstrates the following.
One can see from Fig.~5 that the theory fits the experimental data
well. The theoretical curves in terms of the $W$ function shown in
Figs.~7 and~8 correlate with  experimental $\Delta
T_c(d_\mathrm{Fe2})$ and $\Delta T_c(d_\mathrm{Fe1})$
dependencies. However, in contrast to the case of Fig.~5 where the
experimental data evidence obviously the improvement of the
quality of the S/F interface, the experimental results shown in
Figs. 7 and 8 do not indicate such an improvement. In fact this is
not surprising. The $T_c$-suppression in the S/F structures is the
first order effect in contrast to the case of other small effects
like the superconducting spin valve effect.

Since the experimental results concern the S/F low-quality interface
(Fe/Pb) and the S/F high-quality interface (Fe/Cu/Pb) we will mainly
focus on two simulation curves with the low and high S/F interface
quality, respectively. The analytical results for $W$ are very bulky
so we will not depict them in this paper. Calculations of the
microscopic parameters  $\gamma_b=5.2$ and 2.7 which were obtained
using the results in Fig.~5 fit experimental data satisfactory. All
parameters of the theory were obtained from the normal-state
transport properties of the Pb and Fe layers ($\xi_S$=42 nm,
$l_S$=17 nm, $\xi_F$=7.5 nm and $l_F$=1 nm) and using the results in
Fig.~5 ($\gamma_b$=5.2 and 2.7). Having compared all experimental
data shown in Figs.~7, 8, 11 and 12 with the theoretical results we
can conclude that the {\it general tendencies} of the experimental
dependences are reproduced by the theory. According to theoretical
calculations shown in Fig.~11, the degradation of the quality of the
interface and therefore an increase of $\gamma_b$ leads to the
reduction of the triplet spin valve effect amplitude. This means
that the introduction of the Cu intermediate layer between the Pb
and Fe layers not only stabilizes the S/F interface but also
increases its quality. This conclusion is further confirmed by the
results depicted in Fig.~5. For the bilayer Fe/Pb systems the
introduction of a thin Cu (1.2 nm) film between the Fe and Pb layers
leads to a dramatic drop of the $T_c$. This difference can be easily
described by an improvement of the F/S interface transparency which
is expressed by decreasing the $\gamma_b$ value from 5.2 down to
2.7.

\section{Summary}

By introducing an intermediate thin Cu layer between the
ferromagnetic Fe2 and the superconducting Pb layer in the
superconducting  spin valve heterostructure
CoO$_x$/Fe1/Cu/Fe2/Cu/Pb we have substantially improved its
stability and operational performance. The crucial role of the
intermediate Cu layer is that it prevents material interdiffusion
process and increases electron transparency between the Fe2 and Pb
layers as compared to the previously studied CoO$_x$/Fe1/Cu/Fe2/Pb
heterostructure. Such an improvement has made possible a
comprehensive study of the physical properties of the spin valve
and performing a microscopic analysis of the data beyond the
limitations of the previous studies. All microscopic parameters
were obtained from the normal state transport properties and from
the values of the critical thickness of the S layer. The obtained
realistic parameters of the high quality spin valve
heterostructures enable to reproduce theoretically both the
ordinary spin valve effect and the effect of the generation of the
long-range triplet component of the superconducting condensate in
agreement with our experimental observations. In particular, our
results validate the theoretical description of the long-range
triplet component of the superconducting condensate on a
qualitative level.

\acknowledgments

This work is supported by DFG (Grant. No. LE 3270/1-1). It  was also
partially supported by RFBR (Grants No. 13-02-01389-a and
14-02-00350-a), by Programs of the RAS, by the Ministry of Education
and Science of the Russian Federation, and the program ``5top100''.

\end{document}